\documentclass[letterpaper, 12pt]{article}

\usepackage{algorithm}
\usepackage{algpseudocode}
\usepackage{amsmath}
\usepackage{multirow}
\usepackage{pgfplots}
\pgfplotsset{compat=1.18}
\usepackage{xcolor} 

\usepackage[T1]{fontenc}

\usepackage{geometry}
\usepackage{setspace}

\usepackage[style = chem-acs, articletitle = true]{biblatex}
\usepackage{graphicx}
\usepackage{float}
\newfloat{scheme}{htbp}{los}
\floatname{scheme}{Scheme}
\floatname{chart}{Chart}
\newfloat{graph}{htbp}{loh}

\usepackage{chemformula}
\usepackage[version = 4]{mhchem} 
\usepackage{subcaption}

\usepackage{authblk}
\author[1]{Abir Haque*}
\affil[1]{College of Computing, Georgia Institute of Technology, Atlanta, GA, USA}
\author[1]{Edmond Chow}
\author[2]{Shikhar Shah}
\affil[2]{Department of Mathematics, Emory University, Atlanta, GA, USA}
\author[3]{Andrew J. Medford}
\affil[3]{College of Engineering, Georgia Institute of Technology, Atlanta, GA, USA}
\author[4]{John E. Pask}
\affil[4]{Physics Division, Lawrence Livermore National Laboratory, Livermore, CA, USA}
\author[1,3]{Phanish Suryanarayana}

\title{Highly Parallel Real-Space Random Phase Approximation Using Lanczos Quadrature and Interpolation}
\date{*Email: abirhaque@gatech.edu}

\begin{document}
\maketitle

\begin{abstract}
We present a highly parallelizable, matrix-free, real-space method for computing the random phase approximation (RPA) correlation energy within Kohn-Sham density functional theory. In particular, we avoid the explicit construction and eigen-decomposition of the response function matrix and use Lanczos quadrature to evaluate the trace of a matrix function on a real-space grid. We also show it is possible to exploit the spatial smoothness of the RPA correlation energy density in real-space to reduce computational cost. Specifically, we compute the energy density on a coarse grid, then reconstruct an approximation to the full, fine grid via interpolation. We implement this formulation within the SPARC electronic structure package and demonstrate its convergence, accuracy, agreement with planewave results, and scaling. Interpolation can enable an \(8\times\) reduction in the computational prefactor. Given the embarrassingly parallel nature of the proposed method, we achieve near-ideal speedups and near-cubic scaling, thus allowing us to compute the RPA correlation energy for a silicon system with 256 valence electrons at chemical accuracy in less than 30 minutes on 4,096 CPU cores.
\end{abstract}

\section{Introduction}
Electronic structure calculations, particularly those based on Kohn-Sham density functional theory (DFT) \cite{Kohn1965-sf,Hohenberg1964-qx}, are widely used in chemistry and materials science because they offer an excellent balance between computational cost and accuracy. Nevertheless, the computational cost of solving the generalized Kohn-Sham equations generally grows at least cubically with system size, limiting the range of systems that can be studied. This limitation is particularly severe for nonlocal exchange-correlation approximations, whose evaluation requires information beyond the electron density and its derivatives, thereby increasing both algorithmic complexity and computational expense relative to local/semilocal functionals. Among such approaches, the random phase approximation (RPA) is among the most accurate approximations to the exchange-correlation functional in condensed matter systems \cite{Perdew2001-lw}. It includes exact exchange and captures correlation through the spatially nonlocal, frequency-dependent density response within the adiabatic-connection fluctuation-dissipation framework, rather than relying on a local/semilocal approximation to the correlation hole. Consequently, RPA can provide benchmark results due to its ability to accurately describe van der Waals interactions, small-gap systems, and metals \cite{Ren2012-vl,Eshuis2012-lg}.

Conventional RPA formulations, however, remain impractical for even moderate-sized calculations because of their quartic scaling and substantial prefactor \cite{Gonze2016-pr,Enkovaara2010-xo,Kresse1996-oo}. This cost arises primarily from the explicit dependence on both occupied and unoccupied states: construction of the independent-particle density response function matrix requires all unoccupied states, while evaluation of the correlation energy additionally entails numerical integration over frequency. A variety of methods and software implementations have been developed to perform RPA calculations \cite{Rocca2014-wj,Del_Ben2013-kg,Kaltak2014-ln,Nguyen2014-av,Weinberg2024-ru,Lu2017-wq,Del_Ben2015-kn}, but each imposes modeling restrictions or computational bottlenecks that limit its applicability. In particular, most of these approaches are formulated within a planewave basis, which inherently restricts the boundary conditions to be periodic \cite{Martin2020-em}, and which require additional approximations for isolated systems or partially periodic systems. Among specific implementations, BerkeleyGW \cite{BerkeleyGW} and VASP \cite{Kaltak2014-ln} require the precalculation of both occupied and unoccupied states. WEST reduces this requirement to occupied orbitals alone, but still relies on global Fourier transforms, which have limited scalability on distributed architectures because of the extensive inter-process communication required for data redistribution \cite{WEST,9835388,Ayala2021-wj,9652837}. ABINIT instead computes the RPA correlation energy through the conventional approach of eigendecomposing an explicitly constructed response function matrix \cite{Gonze2016-pr}, thereby inheriting the same quartic-scaling limitations described above. Overall, developing sub-quartic-scaling RPA formulations with a low prefactor, minimal global communication, and flexible boundary-conditions remains critical to simulating large, complex chemical systems with chemical accuracy.

Unlike planewave-based approaches, real-space methods are not restricted to periodic boundary conditions and can instead naturally accommodate Dirichlet, Neumann, and mixed boundary conditions, allowing isolated, semi-infinite, and finite systems to be described directly without artificial periodic replication or vacuum padding \cite{Saad2010-kc,Xu2021-he,Zhang2024-vi,Jing2024-yn,kronik2006parsec,castro2006octopus,motamarri2020dft,beck2009real,bhowmik2026bulk}. Such flexibility is especially valuable for surfaces, nanostructures, and charged or polar systems, where periodicity can introduce spurious electrostatic interactions between periodic images. Real-space discretizations are also well suited to high-performance computing, as their inherent locality supports efficient large-scale parallelism, in contrast to the dense operators and global fast Fourier transform (FFT) communication of planewave methods. Within the context of RPA, Shah et al.\ \cite{SC_RPA} proposed a cubic-scaling real-space formulation that avoids explicit construction of the response function matrix, employing a short-term recurrence block Krylov subspace solver combined with subspace iteration and polynomial filtering. Similarly, Zhang et al.\ \cite{ZHANG_SQ} formulated a cubic-scaling real-space method that uses low-rank approximations to avoid explicit construction of the response function matrix. Although both methods achieve cubic scaling and support complex boundary conditions, they are not massively parallelizable, as their eigensolvers rely on orthogonalization and subspace projection steps that require frequent global communication. Perhaps more importantly, both methods truncate the spectrum of the density response function for computational efficiency, an approximation that can be inaccurate for systems whose spectra exhibit long tails.

We present a highly parallelizable, matrix-free calculation method for the RPA correlation energy, where Lanczos quadrature is utilized to directly evaluate the trace of the matrix function on a real-space grid. Specifically, Lanczos quadrature computes individual diagonal entries of the matrix function, where each entry represents the correlation energy density at a given real-space grid point. This results in the method being embarrassingly parallel up to the total number of real-space grid points. To significantly reduce the overall prefactor, we exploit the spatial smoothness of these correlation energy densities. Within this framework, Lanczos quadrature is applied only to a coarse set of grid points, and we subsequently recover approximations for the full fine grid via interpolation. We also show that error cancellation in the global sum of the interpolated correlation energy densities helps achieve chemical accuracy in cases where point-wise errors may be high. Furthermore, we show that the proposed calculation method can achieve up to a \(100\times\) speedup compared to ABINIT, a well-established planewave code, on comparable systems.

The remainder of this paper is organized as follows. First, we present the theoretical foundations of computing the RPA correlation energy in real-space via Lanczos quadrature, with and without interpolation of diagonal entries. Then, we present numerical experiments that discuss the accuracy, scalability, and limitations of our method, which we implement within SPARC, a real-space finite-difference electronic structure software package \cite{Xu2021-he,Zhang2024-vi}. Finally, we present our conclusions and future directions for our proposed framework for RPA. Note that atomic units are used throughout unless otherwise specified.

\section{Background}
The RPA correlation energy is written as \cite{PhysRevA.68.032507}:
\begin{equation}
E_c=\frac{1}{2\pi}\int_0^\infty \text{Tr}[\log(I-\nu\chi_0(i\omega))+\nu\chi_0(i\omega)]\ d\omega, \label{eq:RPA_full}
\end{equation}
where \(\nu=4\pi(\nabla^2)^{-1}\) is the Coulomb operator and \(\chi_0(i\omega)\) is the non-interacting Kohn-Sham density response function at an imaginary frequency \(i\omega\). Neglecting spin and Brillouin zone integration, the response function matrix is explicitly written as \cite{PhysRev.126.413,PhysRev.129.62}:

\begin{equation}
\chi_0(r,r',i\omega)=2\sum^{N_d}_{j,k=1}(f_j-f_k)\frac{\psi^*_j(r)\psi_k(r)\psi^*_k(r')\psi_j(r')}{\varepsilon_j-\varepsilon_k-i\omega},\label{eq:response_function}
\end{equation}
where \(r,r'\) are real-space coordinates, \(N_d\) is the total number of finite difference points in a given computational domain, \(^*\) denotes complex conjugation, and \(f_i\) are the occupations with corresponding eigenpairs \(\psi_i,\varepsilon_i\) of the Hamiltonian \(H\), satisfying:
\begin{equation}
H\psi_i=\varepsilon_i\psi_i.\label{eq:hamiltonian}
\end{equation} Henceforth, we will consider these operators as discretized within the real-space finite difference approach, wherein the above expressions remain unchanged.
To avoid direct calculations of \(\chi_0(i\omega)\), a matrix-free approach involving matrix-vector operations between \(\chi_0(i\omega)\) and a vector \(u\) can be utilized. In particular, one may compute \(r=\chi_0(i\omega)u\) by utilizing the density functional perturbation theory (DFPT) framework \cite{WEST,Schlipf2020-zc,Goncharov2014-sk}. This involves solving the Sternheimer equation for \(N_s\) occupied states:
\begin{equation}
(H-\varepsilon_nI-i\omega I)\Delta\psi_n=-\psi_n\odot u, \quad n=1,\dots, N_s, \label{eq:sternheimer}
\end{equation}
which contains a complex symmetric linear system with \(\odot \) as the Hadamard product. Once \(\Delta\psi_n\) for each occupied state is obtained, \(r\) is calculated by accumulating all corresponding perturbations in electron density across all states.

\section{Trace Approximations}
\subsection{Lanczos Quadrature}
Our primary contribution lies in utilizing interpolation and Lanczos quadrature to approximate the trace of the function of a matrix.
To achieve this, we first compute a subset of the diagonal entries of the matrix function.
While computing all elements of the diagonal of a matrix via Lanczos quadrature, given an efficient matrix-vector product, is a well-established technique within linear algebra and computational chemistry \cite{SHAH_DISSERTATION,SC_RPA,doi:10.1137/16M1104974,10.1021/acs.jctc.1c01291, White2021-dg}, we show how it can be used with interpolation.
Alternative approaches to approximating the trace, such as stochastic Lanczos quadrature \cite{doi:10.1137/16M1104974}, are available, but we do not utilize them in this work due to their reliance on random sampling, which introduces statistical variance. Additionally, low-rank approximations to the trace truncate the spectrum, which may be inaccurate for spectra with long tails. In this section, we discuss the approximation techniques utilized within the proposed method.
Generally, one can directly approximate \(e_j^Tf(A)e_j\) in the expression \begin{equation}
\text{Tr}[f(A)]=\sum^n_{j=1}e^T_jf(A)e_j\label{eq:trace}
\end{equation}
via Lanczos quadrature, where \(e_j\) is the \(j\)th standard basis vector.
Each term \(e_j^Tf(A)e_j\) is approximated via \(m\) steps of the Lanczos algorithm for symmetric \(A\) and starting vector \(e_j\), where \(A\) is typically used in a matrix-free manner via an efficient matrix-vector multiplication.
The Lanczos algorithm then returns a matrix \(V_m\) with orthonormal columns and an \(m\times m\) tridiagonal matrix \(T_m\) satisfying \begin{equation}
AV_m=V_mT_m.\label{eq:Lanczos_decomp}
\end{equation}
This yields the approximation:
\begin{equation}
e_j^Tf(A)e_j\approx[f(T_m)]_{1,1}\label{eq:Lanczos_trace_eval}
\end{equation}
where \(f(T_m)\) is quickly computed via an eigenvalue decomposition since \(m\) is small.

We are interested in the spectrum of \(\nu\chi_0(i\omega)\), a non-symmetric matrix. One may apply a similarity transform on \(\nu\chi_0(i\omega)\) to obtain \(\nu^{1/2} \chi_0(i\omega) \nu^{1/2}\), a symmetric matrix that shares the same spectrum as \(\nu\chi_0(i\omega)\). This allows us to utilize the Lanczos method for quadrature, which has significantly reduced computational complexity compared to the Arnoldi method for non-symmetric problems.
Denote \(f(A):= \log(I-A)+A\) and \(\tilde{\chi}_0(i\omega):=\nu^{1/2} \chi_0(i\omega) \nu^{1/2}\). Equation \ref{eq:RPA_full} can then be rewritten as: 
\begin{equation}
E_c=\frac{1}{2\pi}\int_0^\infty \sum^{N_d}_{j=1}e^T_jf(\tilde{\chi}_0(i\omega))e_j\ d\omega. \label{eq:RPA_substitution}
\end{equation}
We approximate the outer integral via Gauss-Legendre quadrature:
\begin{equation}
E_c\approx\frac{1}{2\pi}\sum_{k=1}^{N_\omega}w_k\sum^{N_d}_{j=1}e^T_jf(\tilde{\chi}_0(i\omega_k))e_j,\label{eq:RPA_approx}
\end{equation}
where \(N_\omega\) is the frequency quadrature order, \(\omega_k\) are the frequency quadrature nodes, and \(w_k\) are the frequency quadrature weights.

\subsection{Interpolating Diagonal Entries}
The computational cost of the proposed method can be significantly reduced by exploiting the spatial smoothness in the diagonal entries of \(f(\tilde{\chi}_0(i\omega))\) on the real-space grid. By evaluating these entries on a coarse grid and then approximating values on the fine grid via an interpolation method, it may be possible to achieve a reduction in the total number of direct Lanczos calls by factors up to \(8\) in 3D bulk materials while maintaining chemical accuracy.
In this paper, we approximately reconstruct the fine grid via trilinear, tricubic, and Fourier interpolation.
We now formally unify the traditional approach to approximating the RPA correlation energy, as seen in Equation \ref{eq:RPA_approx}, with interpolation on a coarse grid.
Define \begin{equation}
d_j(i\omega):= e_j^Tf(T_m)e_j,
\end{equation} where \(T_m\) is the tridiagonal matrix associated with the \(j^{\text{th}}\) grid point. Furthermore, let \(C_g\) denote a coarse subset of grid points parametrized by the coarsening factor \(g\), where \(g\) denotes the spacing between sampled grid points and \(g=1\) corresponds to the full grid.
Let \(I_g\) denote an interpolation operator that approximately reconstructs fine-grid values from the coarse-grid evaluations \(d_{C_g}(i\omega)\).
The interpolated diagonal entries are then given by \begin{equation}\hat{d}(i\omega)=I_g(d_{C_g}(i\omega)),\end{equation} which yields the following approximation \begin{equation} E_c^{(g)}:=\frac{1}{2\pi}\sum_{k=1}^{N_\omega}w_k\sum^{N_d}_{j=1}\hat{d}_j(i\omega_k).\end{equation} In the limiting case \(g=1\), the coarse and fine grid points are identical, and \(I_g\) simply becomes the identity operator, implying that \(\hat{d_j}=d_j\) for all grid points.
As a result, \(E_c^{(g)}\) reduces exactly to the right-hand side of Equation \ref{eq:RPA_approx}.
Therefore, the proposed RPA formulation with interpolation is simply a generalization of standard Lanczos quadrature, with \(g\) as the parameter that influences computational cost and accuracy via grid coarsening.
For example, an approximately \(8\times\) computational cost reduction is obtained when \(g=2\).

Trilinear interpolation is a local method that provides \(C^0\) continuity by utilizing 2 points along each dimension to linearly approximate the original fine grid points.
To enforce higher-order continuity, we consider tricubic and Fourier interpolation.
Tricubic interpolation, also a local method, provides \(C^2\) continuity by utilizing 4 points along each dimension via cubic polynomials.
When there are periodic boundary conditions, manual wrapping across the domain is necessary for both trilinear and tricubic interpolation.
Fourier interpolation is a global method and assumes the underlying function is periodic. Additionally, Fourier interpolation provides \(C^\infty\) continuity.
Here, we transform the periodic coarse grid values into the frequency domain via a Fourier transform, pad the spectrum to match the fine grid dimensions, then perform an inverse Fourier transform to approximate the fine grid values.
We compare the accuracy of applying trilinear, tricubic, and Fourier interpolation in subsequent numerical experiments.

\subsection{Algorithmic Analysis}\label{alg_analysis}
Regarding serial complexity, we consider \(N_\omega\), the number of frequencies, to be a small constant.
For each frequency \(\omega_k\), we must perform a Lanczos procedure for all \(N_d\) finite difference points.
Each Lanczos procedure will be bounded by \(N_l\) iterations, which will yield an estimate of a given diagonal entry for a given frequency's trace calculation.
Each Lanczos iteration requires a separate loop through \(N_s\) occupied states for each implicit matrix-vector product operation.
For each occupied state, we must solve the Sternheimer equation, which has \(O(N_d)\) complexity assuming the number of iterations of the Krylov subspace method, e.g., Conjugate Orthogonal Conjugate Gradient (COCG), is constant.
Therefore, the complexity of each Lanczos iteration is \(O(N_sN_d)\). This results in each diagonal entry estimate for trace computations requiring a computational complexity of \(O(N_lN_sN_d)\).
Factoring in the loop across \(N_d\) finite difference points gives the overall complexity of our RPA calculation as \(O(N_lN_sN_d^2) \sim O(N_lN_d^3)\).
In practice, \(N_l\) is very small (e.g., set to a maximum of 7 iterations) and independent of system size, hence we treat it as a small prefactor, thus resulting in the final overall serial scaling \(O(N_d^3)\).
Regarding parallel complexity, denote \(p\) as the number of processors.
Each diagonal entry estimation for the trace (i.e., each instance of the Lanczos procedure) requires no communication.
Gathering trace contributions occurs only once per \(\omega_k\), and therefore is treated as having negligible cost.
The parallel complexity simply becomes \(O\left(N_d^3/p\right)\). As \(p\) approaches \(N_d\), the effective complexity becomes \(O(N_d^2)\), with the solution time for the Sternheimer equations to compute the matrix-vector product within the Lanczos procedure being the effective time-to-solution for our RPA formulation on massively parallel systems. Regarding the proposed method's space complexity, the primary terms that dominate storage are the Krylov basis vectors generated during the Lanczos process and the Kohn-Sham orbitals, which scale as \(O(N_lN_d)\) and \(O(N_sN_d)\), respectively. Therefore, the overall space complexity is \(O((N_s+N_l)N_d)\), where \(N_s\) is much larger than \(N_l\). The effective space complexity is \(O(N_sN_d) \sim O(N_d^2)\). The pseudocode for the previously described formulation for calculating the RPA correlation energy is shown in Algorithm \ref{alg:RPA_Lanczos}.
\begin{algorithm}
    \caption{RPA Correlation Energy - Lanczos Quadrature with Interpolation} 
    \label{alg:RPA_Lanczos}
    \begin{algorithmic}[1]
    \State $E_{c}=0$
    \For{$m = 1$ \textbf{to} $N_\omega$}  \Comment{\textcolor{gray}{Loop over frequencies}}
        \State $E^{\text{coarse}}_{c,m}=\{\}$
        \For{$j = 1$ \textbf{to} $N_d$} \Comment{\textcolor{gray}{Loop over grid points}}
            \If{$j$ \text{maps to a coarse grid point}}
                \State $q_{1}=e_j$ \Comment{\textcolor{gray}{Initialize Lanczos 
with \(j\)-th basis vector}}
                \State $\beta_0=0$ 
                \For{$k=1$ \textbf{to} $N_l$} \Comment{\textcolor{gray}{Perform Lanczos iterations}}
                    \State \(r=0\) \Comment{\textcolor{gray}{Apply \(\nu^{1/2} \chi_0(i\omega) \nu^{1/2} q_k\) in matrix-free manner}}

                   \State \(u=\nu^{1/2}q_k\) \Comment{\textcolor{gray}{via Kronecker product }}\Comment{\textcolor{gray}{\(O(N_d^{4/3})\)}}
                    \For{$n=1$ \textbf{to} $N_s$} 
                        \State $(H-\varepsilon_nI-i\omega I)\Delta\psi_n=-\psi_n\odot u$ \Comment{\textcolor{gray}{Solve via COCG} }\Comment{\textcolor{gray}{\(O(N_d)\)}}

  \State $r=r+4\psi_n\odot\text{Re}\{\Delta \psi_n\}$  \Comment{\textcolor{gray}{\(O(N_d)\)}}
                    \EndFor 
                    \State \(v=\nu^{1/2}r\) \Comment{\textcolor{gray}{Complete \(Aq_k\), resume normal Lanczos iteration}}\Comment{\textcolor{gray}{ \(O(N_d^{4/3})\)}}
                    \State $\alpha_k=q_k^T v$
                    \State $v=v-\alpha_k 
q_k-\beta_{k-1}q_{k-1}$
  
                   \If{$k<N_l$}
                        \State $\beta_k=\|v\|_2$
                        \State $q_{k+1}=v/\beta_k$
                    \EndIf

     \EndFor
            \State Construct tridiagonal $T$ from $\alpha$, $\beta$
            \State $TQ = Q \Lambda $
            \State $\text{Append } \sum_{i=1}^{N_l}Q_{1,i}^2\left( \log(1 - \lambda_i) + \lambda_i \right) \text{ to } E^{\text{coarse}}_{c,m}$ \Comment{\textcolor{gray}{Evaluate $e_1^T f(T) e_1$}}
            \EndIf
        \EndFor
    \State \(E_{c,m}=\sum_{i=1}^{N_d}\)\textbf{interpolate(}\(E^{\text{coarse}}_{c,m}\)\textbf{)\(_i\)}
   
 \State $E_c=E_c+E_{c,m}\cdot w_m$

    \EndFor
    \State \textbf{return} \(E_c\)
    \end{algorithmic}
\end{algorithm}

\section{Implementation}
We implement the proposed RPA formulation outlined in Algorithm \ref{alg:RPA_Lanczos} within the SPARC electronic structure package \cite{Jing2025-ei,Ghosh2017-mc,Xu2021-he,Zhang2024-vi,ZHANG_SQ,SC_RPA}.
Distributed parallelism is achieved via the OpenMPI implementation of the Message Passing Interface (MPI), and basic linear algebra routines are performed via LAPACK \cite{gabriel2004open,laug}. Below, we assume no interpolation.

Coarse-grained concurrency naturally exists in the loop that traverses through all \(N_d\) grid points.
In particular, no communication is involved in this loop as can be seen in Algorithm \ref{alg:RPA_Lanczos}.
Therefore, we simply partition \(N_d\) total grid indices across all \(p\) processors such that each processor \(j'\) owns \(N^{\text{local}}_d\leq\lceil \frac{N_d}{p} \rceil\) indices, and therefore computes \begin{equation} E^{\text{local}}_{c}:=\sum^{(j'+1)N^{\text{local}}_d\ -\ 1}_{j=j'N^{\text{local}}_d}e^T_jf(T_m)e_j,\end{equation} where \(T_m\) is the tridiagonal matrix associated with the \(j^{\text{th}}\) grid point. The electronic correlation energy \(E^{\text{global}}_{c}\) at any given \(\omega_k\) is computed via a single global reduction, e.g., \texttt{MPI\_Reduce}.

The \(\tilde{\chi}_0(i\omega)q\) matrix-vector product that arises within Lanczos quadrature is impractical to perform when utilizing an explicit construction of  \(\tilde{\chi}_0(i\omega)\). Therefore, we compute \(u=\nu^{1/2}q\) via a Kronecker product-based formalism in real-space \cite{Jing2024-yn,Jing2025-ei}. Then, for each occupied Kohn-Sham orbital, we solve Equation \ref{eq:sternheimer} and accumulate the corresponding perturbation in the electron density into \(r\). Finally, we obtain the full matrix-vector result by performing \(v=\nu^{1/2}r\).

We efficiently solve the linear system in Equation \ref{eq:sternheimer} via COCG, with the initial guess constructed via a Galerkin projection \cite{Van_der_Vorst1990-dd,SC_RPA}. In particular, the initial guess effectively deflates the spectrum of Equation \ref{eq:sternheimer}, which permits overall convergence in a few iterations. The complexity of generating this initial guess for each linear system is \(O(N_dN_s)\). However, it is negligible to perform in practice given modest sizes of \(N_s\). Within our parallelization scheme, the solution to Equation \ref{eq:sternheimer} via COCG is considered a fine-grained task. Although the number of iterations required to solve Equation \ref{eq:sternheimer} for a given frequency at each grid point across the domain is uneven, this load imbalance is negligible. In addition, the variance in the number of COCG iterations taken for a given \(\omega_k\) is low, so we do not require advanced load-balancing schemes.

Standard interpolation methods, such as tricubic and Fourier interpolation, are computationally inexpensive on small grid sizes. However, the communication cost across distributed memory would outweigh any theoretical performance benefits from utilizing many processors \cite{Ayala2021-wj}. Therefore, we only perform interpolation on a single processor within our implementation. If interpolation is enabled within SPARC, then the global reduction step is replaced with a collective communication step, e.g., \texttt{MPI\_Gatherv}, to obtain each processor's set of local coarse grid points so the root processor can perform interpolation, and then obtain an approximation to the electronic correlation energy. Most importantly, only one collective communication operation is required within SPARC at each \(\omega_k\), which enables SPARC to scale with nearly ideal speedups for massive numbers of processors across distributed memory.

\section{Numerical Results and Discussion}
We now discuss the accuracy and performance of the Lanczos quadrature-based RPA calculation. We use \(\Gamma\)-point Brillouin zone integration, the Perdew-Burke-Ernzerhof (PBE) \cite{PhysRevLett.77.3865} exchange-correlation functional, and Optimized Norm-Conserving Vanderbilt (ONCV) \cite{PhysRevB.88.085117} pseudopotentials with nonlinear core corrections from the Shojaei-Pask-Medford-Suryanarayana (SPMS) \cite{SHOJAEI2023108594} table. The order of Gauss-Legendre quadrature for the frequency integral is set to \(N_\omega=8\).

\subsection{Chemical Systems}
We consider several systems comprised of cells of silicon (Si), carbon (C), and lithium hydride (LiH) in our numerical experiments.

We consider 2-atom cubic cells with side lengths of 5.14, 3.36, and 4.38 Bohr for Si\(_2\), C\(_2\), and LiH systems, containing 8, 8, and 4 electrons, respectively. For cross-verification with ABINIT, we use results from Zhang et al.\ \cite{ZHANG_SQ} who used all unoccupied orbitals for constructing \(\chi_0\) and utilized planewave cutoffs of 135, 195, and 165 Ha for the Si, C, and LiH systems, respectively. No perturbations were applied to the atoms of these systems.

We also consider Si\(_8\), Si\(_{32}\), Si\(_{64}\) systems where we randomly perturb atom positions such that bond lengths are changed from their resting lengths by up to \(5\%\). Denote \(x_i\) as the position vector of atom \(i\). The perturbations are obtained via:
\begin{equation}
x^{\text{perturbed}}_i=x^{\text{unperturbed}}_i+\delta\cdot r_i,
\end{equation}
where \(\delta\) is \(10\%\) of the distance to the nearest atom in the unperturbed system and \(r_i\) is a random vector drawn uniformly from the interval \([-0.5,0.5]^3\).

Lastly, we examine the performance of the proposed method's implementation within SPARC. We construct larger cells of Si\(_8\), Si\(_{32}\), Si\(_{64}\), (LiH)\(_4\), (LiH)\(_{16}\), and (LiH)\(_{32}\), all of which have perturbed atom positions.

\subsection{Error Metrics}
To evaluate the accuracy of various interpolation methods for computing RPA energy, we use two key error metrics that compare the interpolated diagonal entry values against the corresponding baseline values calculated directly.
Recall \(d_i\) is the directly calculated value at grid point \(i\) for some \(\omega_k\) and \(\hat{d_i}\) is the interpolated value at grid point \(i\) for the same \(\omega_k\).
Recall \(E_c=\sum_id_i\) and \(\hat{E_c}=\sum_i \hat{d_i}\). The absolute error between \(E_c\) and \(\hat{E_c}\) is \begin{equation}|\Delta E_c |=|E_c-\hat{E_c}|=\bigg|\sum_i d_i-\sum_i \hat{d_i}\bigg|=\bigg|\sum_i (d_i-\hat{d_i})\bigg|=\bigg|\sum_i \Delta d_i\bigg|, \end{equation} where \( \Delta d_i\) is the signed point-wise error between \(d_i\) and \(\hat{d_i}\).
The \(L_1\) error between \(d\) and \(\hat{d}\), i.e., \begin{equation}\sum_i|d_i-\hat{d_i}|=\sum_i| \Delta d_i|\end{equation} quantifies the total point-wise errors \(\Delta d_i\).
Given that a subset of the point-wise errors \(\Delta d_i\) are positive and the rest are negative (or exactly 0 at coarse grid points), we can immediately see that portions of these errors present in the \(L_1\) error will cancel out in the overall absolute error \(|\sum_i\Delta d_i |\).
Therefore, we quantify the error cancellation ratio (ECR) as the ratio between the overall absolute error and the \(L_1\) error, i.e., \begin{equation}\text{ECR}=\frac{\bigg|\sum_i\Delta d_i \bigg|}{\sum_i|
\Delta d_i|}.\end{equation} An ECR that is small means that a given interpolation method that computed \(\hat{d_i}\) benefits from nearly-perfect error cancellation. However, a small ECR must be interpreted carefully in the context of the overall accuracy. If the \(L_1\) error is exceptionally low, then \(\hat{d}_i\) generally fits very well with \(d_i\); therefore not much error can be canceled with respect to less accurate methods. Such cases where the \(L_1\) error is low include when the underlying function is smooth.

\subsection{Convergence with Respect to Parameters}

\begin{figure}[H]
\centering

\begin{subfigure}[b]{0.48\textwidth}
\centering
\begin{tikzpicture}
\begin{axis}[
height=5.5cm,
width=\linewidth, 
xlabel=Grid Spacing,
ylabel=Error,
x dir=reverse,
font=\footnotesize, 
ymode=log,
xmode=log,
grid=major,
ymin=0.000005, 
ymax=0.1,
xmin=0.09,
xtick={0.6,0.5,0.4,0.3,0.2,0.1},
xticklabels={0.6,0.5,0.4,0.3,0.2,0.1},
ytick={0.1,0.01, 0.001, 0.0001, 0.00001, 0.000001},
yticklabels={$10^{-1}$,$10^{-2}$, $10^{-3}$, $10^{-4}$, $10^{-5}$}, 
legend style={
        at={(0.5,-0.3)}, 
        anchor=north,    
        legend columns=3,
    },
]
\addplot[mark=square*,color=teal] 
coordinates {
(0.197829,	1.1072E-05)
(2.86E-01,	4.3059E-05)
(5.14E-01,	0.000349402)
(6.43E-01,	0.002075371)

};
\addlegendentry{Si\(_2\), \(g=1\)}

\addplot[mark=*, color=blue] coordinates {
(0.560737,	0.00346305)
(0.420553,	0.000338482)
(0.336443,	0.000104564)
(0.210277,	0.000058163)
(0.140184,	2.2138E-05)
};
\addlegendentry{C\(_2\), \(g=1\)}

\addplot[mark=triangle*, color=red] coordinates {
(0.136969,	1.87141E-05)
(0.273938,	4.30669E-05)
(0.36525	,0.000145778)
(0.4383	,0.000622532)
(0.547875,	0.002309215)

};
\addlegendentry{LiH, \(g=1\)}

\addplot[mark=square,mark options={solid}, color=teal, mark size=4pt,very thick,dotted] coordinates {
(0.197829,	1.108E-05)
(2.86E-01,	4.3006E-05)
(5.14E-01,	0.000325362)
(6.43E-01,	0.005234301)

};
\addlegendentry{Si\(_2\), \(g=2\)}

\addplot[mark=o,mark options={solid}, color=blue, mark size=4pt,very thick,dotted] coordinates {
(0.560737,	0.014780719)
(0.420553,	0.011808196)
(0.336443,	0.000189966)
(0.210277,	0.000140916)
(0.140184,	2.242E-05)

};
\addlegendentry{C\(_2\), \(g=2\)}

\addplot[mark=triangle,mark options={solid}, color=red, mark size=5pt,very thick,dotted] coordinates {

(0.136969,	1.77207E-05)
(0.273938,	0.000259637)
(0.36525	,0.003662909)
(0.4383,	0.013902781)
(0.547875,	0.068487481)

};
\addlegendentry{LiH, \(g=2\)}

\end{axis}
\end{tikzpicture}
\caption{Grid spacing.}
\label{fig:grid_spacing}
\end{subfigure}
\hfill
\begin{subfigure}[b]{0.48\textwidth}
\centering
\begin{tikzpicture}
\begin{axis}[
height=5.5cm,
width=\linewidth,
xlabel=Sternheimer Tolerance,
ylabel=Error,
x dir=reverse,
font=\footnotesize,  
xmode=log,
ymode=log,
grid=major,
xmin=0.009, 
xmax=0.25, 
ymin=0.0000005, 
ymax=0.01,
ytick={0.01, 0.001, 0.0001, 0.00001, 0.000001, 0.0000001},
yticklabels={$10^{-2}$, $10^{-3}$, $10^{-4}$, $10^{-5}$, $10^{-6}$, $10^{-7}$},
xtick={0.2, 0.1, 0.05, 0.01},
xticklabels={0.2, 0.1, 0.05, 0.01},
legend style={
        at={(0.5,-0.3)},
        anchor=north,   
  legend columns=3,  
    },
]

\addplot[mark=square*,color=teal] coordinates {
(2.00E-01,	0.002376194)
(1.00E-01,	0.000665591)
(5.00E-02,	9.2527E-05)
(1.00E-02,	4.1828E-05)

};
\addlegendentry{Si\(_2\), \(g=1\)}

\addplot[mark=*, color=blue] coordinates {
(2.00E-01,	0.00209376)
(1.00E-01,	0.001032425)
(5.00E-02,	8.4587E-05)
(1.00E-02,	8.662E-06)

};
\addlegendentry{C\(_2\), \(g=1\)}

\addplot[mark=triangle*,color=red] coordinates {
(2.00E-01,	0.000914898)
(1.00E-01,	0.000337612)
(5.00E-02,	0.000101011)
(1.00E-02,	7.3593E-06)

};
\addlegendentry{LiH, \(g=1\)}

\addplot[mark=square,mark options={solid}, color=teal, mark size=4pt,very thick,dotted] coordinates {
(2.00E-01,	0.002363359)
(1.00E-01,	0.000671578)
(5.00E-02,	9.3759E-05)
(1.00E-02,	4.4518E-05)

};
\addlegendentry{Si\(_2\), \(g=2\)}

\addplot[mark=o,mark options={solid}, color=blue, mark size=4pt,very thick,dotted] coordinates {
(2.00E-01,	0.002092868)
(1.00E-01,	0.001030252)
(5.00E-02,	8.4676E-05)
(1.00E-02,	8.638E-06)

};
\addlegendentry{C\(_2\), \(g=2\)}

\addplot[mark=triangle,mark options={solid}, color=red, mark size=5pt,very thick,dotted] coordinates {

(2.00E-01,	0.000913879)
(1.00E-01,	0.000335905)
(5.00E-02,	0.000100938)
(1.00E-02,	7.3734E-06)

};
\addlegendentry{LiH, \(g=2\)}

\end{axis}
\end{tikzpicture}
\caption{Sternheimer tolerance.}
\label{fig:stern_tol}
\end{subfigure}

\vspace{0.5cm} 
\begin{subfigure}[b]{0.48\textwidth}
\centering
\begin{tikzpicture}
\begin{axis}[
height=5.5cm,
width=\linewidth,
xlabel=Lanczos Iterations, 
ylabel=Error,
font=\footnotesize, 
ymode=log,
grid=major,
ymin=0.0000000005, 
ymax=0.01,
ytick={0.001, 0.00001, 0.0000001, 0.000000001},
yticklabels={ $10^{-3}$,$10^{-5}$, $10^{-7}$, $10^{-9}$},
xmin=1.8, 
xmax=4.2,
xtick={2,3,4},
xticklabels={2,3,4},
legend style={
        at={(0.5,-0.3)},   
        anchor=north,      
        legend columns=3,  
    },
]

\addplot[mark=square*,color=teal] coordinates {
(2,	0.002568228)
(3,	7.2312E-05)
(4,	1.824E-06)

};
\addlegendentry{Si\(_2\), \(g=1\)}

\addplot[mark=*, color=blue] coordinates {
(2,	0.005489079)
(3,	9.0098E-05)
(4,	8.16E-07)

};
\addlegendentry{C\(_2\), \(g=1\)}

\addplot[mark=triangle*, color=red] coordinates {
(2,	0.000105463)
(3,	6.636E-07)
(4,	2.8E-09)

};
\addlegendentry{LiH, \(g=1\)}

\addplot[mark=square,mark options={solid}, color=teal, mark size=4pt,very thick,dotted] coordinates {
(2,	0.002565482)
(3,	6.9569E-05)
(4,	9.13E-07)

};
\addlegendentry{Si\(_2\), \(g=2\)}

\addplot[mark=o,mark options={solid}, color=blue, mark size=4pt,very thick,dotted] coordinates {
(2,	0.005489083)
(3,	8.9799E-05)
(4,	8.12E-07)

};
\addlegendentry{C\(_2\), \(g=2\)}

\addplot[mark=triangle,mark options={solid}, color=red, mark size=5pt,very thick,dotted] coordinates {
(2,	0.000105462)
(3,	6.627E-07)
(4,	1.9E-09)

};
\addlegendentry{LiH, \(g=2\)}

\end{axis}
\end{tikzpicture}
\caption{Lanczos iterations.}
\label{fig:lanczos_iterations}
\end{subfigure}
\hfill

\begin{subfigure}[b]{0.48\textwidth}
\centering
\begin{tikzpicture}
\begin{axis}[
height=5.5cm,
width=\linewidth,
xlabel=Grid Coarsening Factor, 
ylabel=Error,
font=\footnotesize, 
x dir=reverse,
ymode=log,
grid=major,
ymin=0.0000000005, 
ymax=0.001,
ytick={0.001, 0.0001, 0.00001, 0.000001, 0.0000001, 0.00000001, 0.000000001},
yticklabels={ $10^{-3}$, $10^{-4}$, $10^{-5}$, $10^{-6}$, $10^{-7}$, $10^{-8}$, $10^{-9}$},
legend style={
        at={(0.5,-0.3)},   
        anchor=north,      
        legend columns=3,  
    },
]

\addplot[mark=square*,color=teal] coordinates {

(2, 1.6e-08)
(4,1.7e-08)
(6,1.3e-04)
};
\addlegendentry{Si\(_2\)}

\addplot[mark=*, color=blue] coordinates {
(2,4.6e-09)
(4,3.8e-07)
(8, 1.5e-04)
};
\addlegendentry{C\(_2\)}

\addplot[mark=triangle*, color=red] coordinates {

(2,8.5e-10)
(4,7.8e-06 )
(8, 1.8e-04)

};
\addlegendentry{LiH}

\end{axis}
\end{tikzpicture}
\caption{Grid coarsening factor.}
\label{fig:grid_coarsening}
\end{subfigure}

\caption{Convergence of the RPA correlation energy (Ha/atom) with respect to various parameters.}
\label{fig:RPA_convergence_all}
\end{figure}
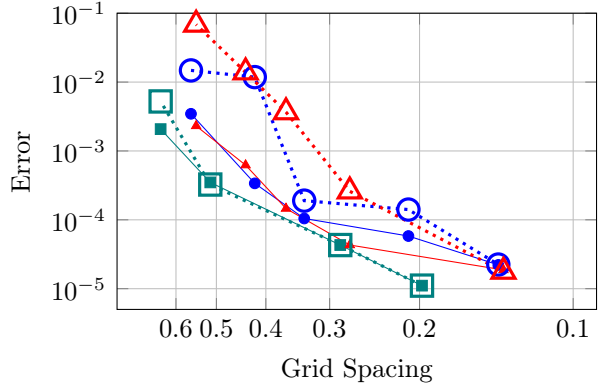
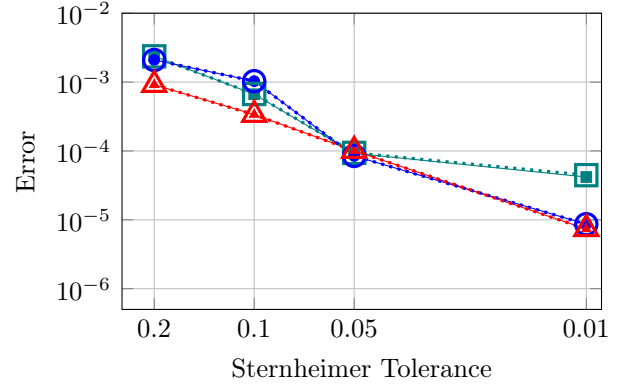
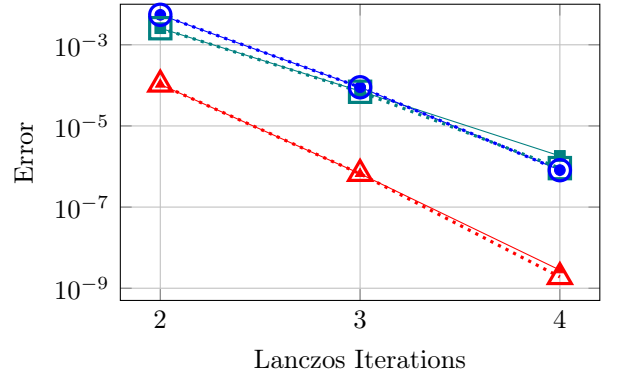
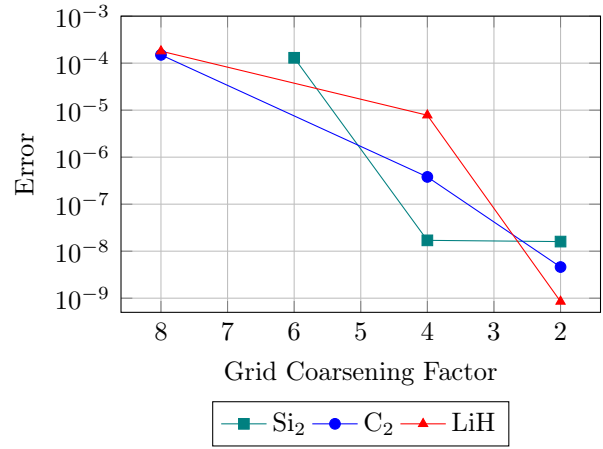

We now consider how the RPA correlation energy converges with respect to the grid spacing of the finite difference discretization, number of Lanczos iterations, grid coarsening, and Sternheimer tolerance, which is the relative residual in the Euclidean norm in approximating the solution to Equation \ref{eq:sternheimer} via COCG.
The absolute errors in Figure \ref{fig:RPA_convergence_all} are defined with baseline calculations using grid spacings \(h\approx\) \(0.07\), \(0.14\), \(0.08\) Bohr for the C\(_2\), Si\(_2\), and LiH systems, respectively.
Furthermore, these baseline calculations use \(N_l=7\), a Sternheimer tolerance of \(10^{-3}\), and a coarse grid spacing of \(g=1\), i.e., no interpolation.

According to Figure \ref{fig:grid_spacing}, we observe rapid convergence in the correlation energy with smaller grid spacings.
Specifically, we observe that the correlation energy rapidly converges to within \(10^{-4}\) Ha/atom as the grid spacing is refined. Furthermore, we see that interpolation may require fine grid spacings versus using \(g=1\) to achieve \(10^{-3}\) Ha/atom convergence for some systems. In Figure \ref{fig:stern_tol}, we see that we consistently achieve correlation energies with errors within \(10^{-4}\) Ha/atom for a Sternheimer tolerance of \(0.05\).
In fact, a loose tolerance of \(0.1\) is sufficient to obtain chemical accuracy of \(\sim 10^{-3}\) Ha/atom. Furthermore, we see that interpolation achieves levels of accuracy that are similar to its full-grid counterpart. We see in Figure \ref{fig:lanczos_iterations} that the RPA correlation energy rapidly converges across subsequent Lanczos iterations, which indicates that \(N_l=4\) Lanczos iterations can be used to safely obtain chemical accuracy. Additionally, we see that interpolation also has negligible effects on convergence. Lastly, Figure \ref{fig:grid_coarsening} shows how increasing the coarsening factor \(g\) on a fine baseline affects chemical accuracy when utilizing Fourier interpolation. We see that small grid coarsening factors, with other parameters set to be very strict, result in consistently small RPA errors. The study in Figure \ref{fig:grid_coarsening} is presented in order to isolate the error in interpolation from other sources of error, and should not be conflated with overall system convergence, which accounts for other errors in the underlying quadrature calculation. The interpolation accuracy is not solely controlled by \(g\). Instead, additional aspects of a given chemical system will influence the interpolation accuracy, such as the interpolation method, \(L_1\) error, error cancellation, and atom positions.

\subsection{Accuracy}
We evaluate the accuracy of the proposed RPA calculation in SPARC by verifying its agreement with ABINIT results reported by Zhang et al.\ \cite{ZHANG_SQ}. Since ABINIT utilizes a planewave basis set versus real-space finite-difference, this comparison serves as a cross-verification of consistency rather than a ground truth comparison. For these evaluations, we use a grid spacing of \(h=0.2, 0.14,\) and \(0.15\) Bohr for the Si, C, and LiH systems, respectively. Lastly, we set \(N_l=4\), a Sternheimer tolerance of \(10^{-2}\), and coarsening factor of \(g=1\). Table \ref{table:accuracy_comparison} shows the accuracy comparison for Si, C, and LiH. In particular, we see that we achieve agreement between SPARC and ABINIT within \(10^{-4}\) Ha/atom. The agreement is expected to further increase if larger planewave cutoffs are chosen in ABINIT.
\begin{table}[H]
\centering
\begin{tabular}{|l|c|c|c|}
\hline
\textbf{System} & \textbf{ABINIT (Ha/atom)} & \textbf{SPARC (Ha/atom)} & \textbf{Energy Difference (Ha/atom)} \\
\hline
Si\(_2\) & -0.20246 & -0.20267 & \(2.1\times10^{-4}\) \\
C\(_2\) &  -0.21899
&  -0.21934&\(3.5\times10^{-4}\) \\
LiH &  -0.06299 &  -0.06340&\(4.1\times10^{-4}\)  \\
\hline
\end{tabular}
\caption{Energy difference (Ha/atom) between ABINIT and SPARC RPA calculations, the latter without interpolation.}
\label{table:accuracy_comparison}
\end{table}

We evaluate the accuracy of various interpolation schemes applied to the matrix diagonals by comparing the difference in the RPA correlation energy approximation in SPARC with and without interpolation. In particular, we present a study of using interpolation for computing the correlation energy for larger chemical systems with perturbations from a perfect crystal structure. As previously observed in Figure \ref{fig:grid_spacing}, not all chemical systems may benefit from interpolation in terms of performance due to significantly less interpolation accuracy obtained when utilizing a given grid-spacing. Hence, we only focus on various Si systems when assessing interpolation accuracy for larger systems. The grid spacing utilized in these series of experiments is \(h\approx0.51\) Bohr. We set \(N_l=3\), a Sternheimer tolerance of \(10^{-1}\), and coarsening factor \(g=2\). Table \ref{table:large_interp_err} summarizes the results for Si\(_8\), Si\(_{32}\), and Si\(_{64}\). Overall, cubic and Fourier interpolation give errors that are consistently smaller than chemical accuracy. In general, trilinear interpolation is inadequate for these systems.

\begin{table}[H]
\centering
\resizebox{\textwidth}{!}{\begin{tabular}{|l|ccc|ccc|ccc|}
\hline
 & \multicolumn{3}{c|}{\textbf{Energy Difference (Ha/atom)}} & \multicolumn{3}{c|}{\textbf{$L_1$ Error}} & \multicolumn{3}{c|}{\textbf{ECR Metric}} \\
\textbf{System} &  \textbf{Trilinear} & \textbf{Tricubic} & \textbf{Fourier} & \textbf{Trilinear} & \textbf{Tricubic} & \textbf{Fourier} & \textbf{Trilinear} & \textbf{Tricubic} & \textbf{Fourier} \\
\hline
Si\(_8\) & \(2.5\times10^{-3}\) & \(7.2\times10^{-4}\)  & \(1.0\times10^{-4}\)  & \(2.1\times10^{-2}\)  & \(8.2\times10^{-3}\)  &\(8.8\times10^{-3}\) & \(11.8\%\)  & \(8.8\%\)  & \(1.2\%\)  \\
Si\(_{32}\)   & \(1.3\times10^{-3}\)  & \(3.4\times10^{-4}\)  & \(7.7\times10^{-5}\)  & \(2.1\times10^{-2}\)  & \(9.1\times10^{-3}\) & \(9.7\times10^{-3}\) & \(6.4\%\) & \(3.7\%\)  &\(0.8\%\) \\
Si\(_{64}\)      & \(1.2\times10^{-3}\)  & \(3.0\times10^{-4}\)  & \(2.7\times10^{-5}\)  & \(2.1\times10^{-2}\)  & \(9.0\times10^{-3}\) & \(9.7\times10^{-3}\) & \(5.8\%\) & \(3.3\%\)  &\(0.3\%\) \\
\hline
\end{tabular}}
\caption{Comparison of the energy difference (Ha/atom) compared to SPARC RPA calculation without interpolation for large, perturbed systems }
\label{table:large_interp_err}
\end{table}

\begin{figure}[h]
\centering
\includegraphics[width=0.6\textwidth]{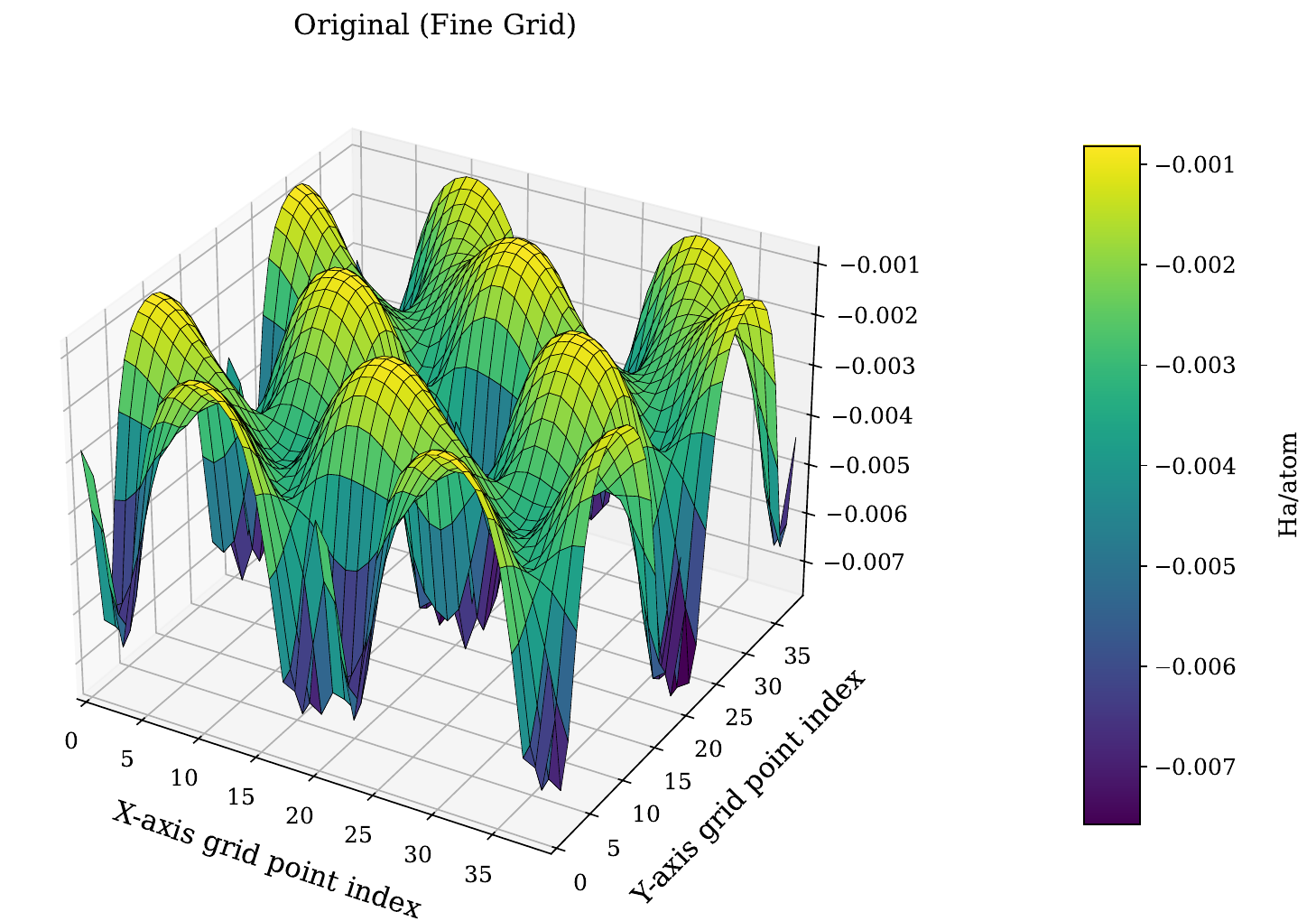}
\caption{Cross section of \(d_i\) with no interpolation for Si\(_{64}\) at \(z=0\) for \(\omega=0.02\), where \(z\in[0,39]\) }
\label{fig:original_cross_section}
\end{figure}

\begin{figure}[!htbp]
\centering
\includegraphics[width=0.96\textwidth]{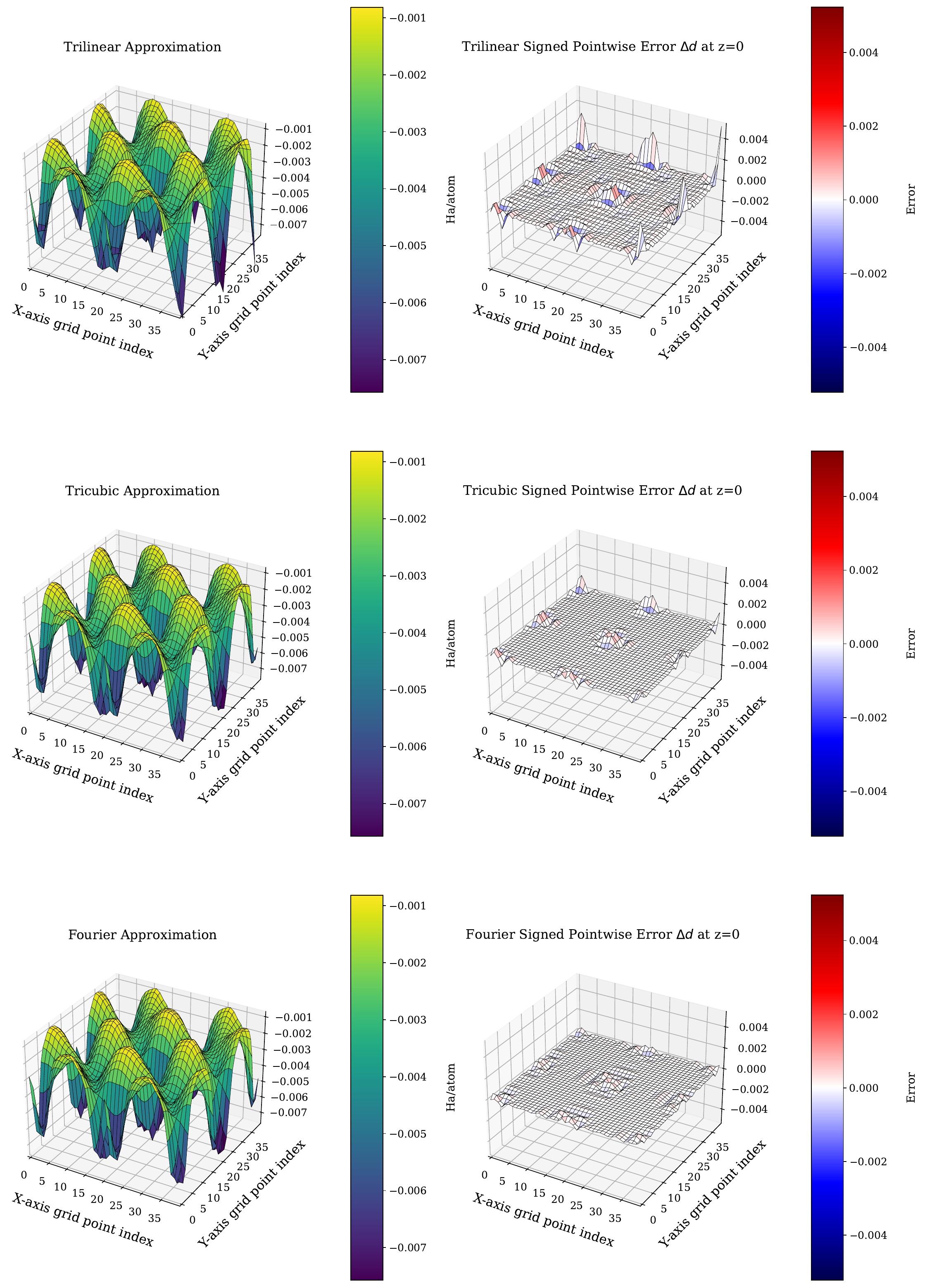}
\caption{Cross sections of interpolated grids \(\hat{d_i}\) for Si\(_{64}\) at \(z=0\) for \(\omega=0.02\), where \(z\in[0,39]\).}
\label{fig:analysis_cross_section}
\end{figure}

\subsection{Error Cancellation}
We empirically analyze the role of error cancellation in computing the correlation energy via interpolation for various chemical systems by comparing the absolute error in the computed energies, the \(L_1\) error, and ECR. We also show cross sections of \(d_i\) and various computations of \(\hat{d_i}\) for Si\(_{64}\) in real-space in Figures \ref{fig:original_cross_section} and \ref{fig:analysis_cross_section}.

According to Table \ref{table:large_interp_err}, we see that Fourier interpolation exhibits slightly higher \(L_1\) errors than tricubic interpolation.
Although both methods achieve chemical accuracy, Fourier interpolation consistently outperforms tricubic interpolation in absolute error. This is likely due to the Fourier method benefiting from error cancellation, as the ECR scores are significantly lower for Fourier than for tricubic.

Chemical systems may observe sharp variations at grid points nearest to atom positions, such as with Si\(_{64}\) in Figure \ref{fig:original_cross_section}. Such sharp variations are likely artifacts that arise from utilizing coarse grids that do not fully capture the local behavior. In such situations, a given interpolation method may struggle to perfectly minimize the point-wise errors globally, as seen in Figure \ref{fig:analysis_cross_section}. This tends to arise when the coarsened grid does not capture a substantial portion of sharp variations in the geometry. However, computing the correlation energy with chemical accuracy via the proposed method does not necessarily require minimal point-wise error globally. Rather, we require that the overall absolute error in the final correlation energy is
minimized. Results from Table \ref{table:large_interp_err} verify that these sparse, point-wise discrepancies are negligible in computing
the final correlation energy through error cancellation.

\subsection{Performance}
We performed two sets of strong scaling experiments to evaluate the performance of the proposed method. The first set of experiments were performed on systems where interpolation has a similar accuracy to no interpolation, i.e., Si\(_{8}\),  Si\(_{32}\), and Si\(_{64}\). The grid spacing utilized in this series of experiments is \(h\approx0.51\) Bohr. We set \(N_l=3\), a Sternheimer tolerance of \(10^{-1}\), and coarsening factor \(g=2\). The second set of experiments were performed without interpolation for (LiH)\(_4\), (LiH)\(_{16}\), and (LiH)\(_{32}\). The grid spacing utilized in this series of experiments is \(h\approx0.47\) Bohr. We set \(N_l=2\) and a Sternheimer tolerance of \(10^{-1}\).

All scaling experiments were performed on the Phoenix supercomputer at the Georgia Institute of Technology. Each compute node has Dual Xeon Gold 6226 CPUs @ 2.7 GHz, 192GB-768GB DDR4-2933 MHz DRAM, and the machine has an Infiniband 100HDR interconnect. We perform scaling tests on up to 4,096 CPU cores distributed across 256 nodes. Fourier interpolation was used in these experiments.

\begin{figure}[htbp]
\centering
\begin{subfigure}[b]{0.48\textwidth}
\centering
\begin{tikzpicture} 
\begin{axis}[
width=\linewidth,
height=7cm,
ymin=1, 
ymax=50000,
ytick={1,10,100,1000, 10000, 100000},
yticklabels={$10^{0}$,$10^{1}$,$10^{2}$,$10^{3}$, $10^{4}$},
xlabel={CPU Cores},
ylabel={Wall Time (Seconds)},
xmode=log, ymode=log,
log basis x={2}, 
xtick={128, 256, 512, 1024, 2048, 4096},
xticklabels={128, 256, 512, 1024, 2048, 4096},
grid=major, 
legend pos=south east, 
]
\addplot[ color=blue, mark=square, ] coordinates { (128, 31.850) (256, 16.899) (512, 9.127) (1024, 4.773) };
\addlegendentry{Si\(_{8}\)}
\addplot[ color=blue, mark=*, ] coordinates { (512, 801.895) (1024, 449.837) (2048, 221.042) (4096, 132.782) };
\addlegendentry{ Si\(_{32}\)}
\addplot[ color=blue, mark=triangle, ] coordinates { (512, 10728.875) (1024, 5086.300) (2048, 2529.191) (4096, 1607.614) };
\addlegendentry{Si\(_{64}\)}
\addplot[ color=gray,dotted,ultra thick ] coordinates { (128, 31.850) (256, 15.925) (512, 7.9625) (1024, 3.98125) };
\addplot[ color=gray,dotted,ultra thick ] coordinates { (512, 801.895) (1024, 400.9475) (2048, 200.47375) (4096, 100.236875) };
\addplot[ color=gray,dotted,ultra thick ] coordinates { (512, 10728.875) (1024, 5364.4375) (2048, 2682.21875) (4096, 1341.109375) };
\end{axis}
\end{tikzpicture}
\caption{Si\(_{8}\), Si\(_{32}\), Si\(_{64}\) wall time.}
\label{fig:Si_scaling}
\end{subfigure}
\begin{subfigure}[b]{0.48\textwidth}
\centering
\begin{tikzpicture} 
\begin{axis}[
width=\linewidth,
height=7cm,
ymin=1, 
ymax=10000,
ytick={1,10,100,1000, 10000},
yticklabels={$10^{0}$,$10^{1}$,$10^{2}$,$10^{3}$},
xlabel={CPU Cores},
ylabel={Wall Time (Seconds)},
xmode=log, ymode=log,
log basis x={2}, 
xtick={128, 256, 512, 1024, 2048, 4096},
xticklabels={128, 256, 512, 1024, 2048, 4096},
grid=major, 
legend pos=south east, 
]
\addplot[ color=blue, mark=square, ] coordinates { (128,10.558) (256,5.554) (512,3.228) };
\addlegendentry{(LiH)\(_{4}\)}
\addplot[ color=blue, mark=*, ] coordinates { (256,489.160) (512,260.360) (1024,126.817) (2048,63.183) };
\addlegendentry{(LiH)\(_{16}\)}
\addplot[ color=blue, mark=triangle, ] coordinates { (512,3235.586) (1024,1623.705) (2048,824.201) };
\addlegendentry{(LiH)\(_{32}\)}
\addplot[ color=gray,dotted,ultra thick ] coordinates { (128,10.558) (256,10.558/2) (512,10.558/4) };
\addplot[ color=gray,dotted,ultra thick ] coordinates { (256,489.160) (512,489.160/2) (1024,489.160/4) (2048,489.160/8) };
\addplot[ color=gray,dotted,ultra thick ] coordinates { (512,3235.586) (1024,3235.586/2) (2048,3235.586/4) };
\end{axis}
\end{tikzpicture}
\caption{(LiH)\(_{4}\), (LiH)\(_{16}\), (LiH)\(_{32}\) wall time.}
\label{fig:LiH_scaling}
\end{subfigure}
\caption{Strong scaling results. \textcolor{gray}{\textbf{Gray}} dotted lines represent ideal scaling.}
\label{fig:combined_scaling}
\end{figure}
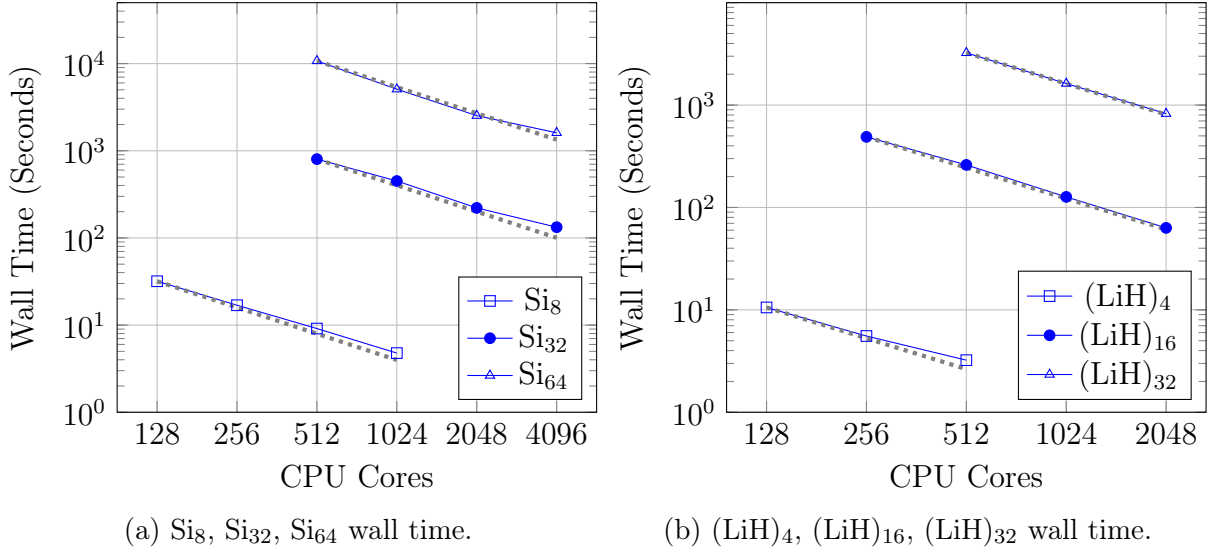

The scaling results presented in Figures \ref{fig:Si_scaling} and \ref{fig:LiH_scaling} show the excellent scalability offered by the proposed RPA calculation within SPARC with interpolation.
In particular, we observe near-ideal speedups.
This is entirely due to the embarrassingly parallel nature of the \(N_d\) loop outlined in Algorithm \ref{alg:RPA_Lanczos}, with and without interpolation, where we evaluate matrix diagonal entries independently via Lanczos quadrature.
Furthermore, Fourier interpolation reduces the runtime without sacrificing chemical accuracy with respect to not utilizing interpolation for the Si systems, as previously seen in Table \ref{table:large_interp_err}. The scaling results also indicate that the proposed RPA calculation has \(O(N_d^{3.38})\) and \(O(N_d^{3.30})\) scaling for the LiH and Si systems, respectively, on 512 cores. As previously mentioned in Section \ref{alg_analysis}, the reason we achieve near-cubic scaling instead of exactly cubic in practice is because we do not utilize a constant number of COCG iterations in solving each Sternheimer equation, which have different diagonal shifts across different values of \(\omega_k\) and \(\varepsilon_n\). The (LiH)\(_4\), (LiH)\(_{16}\), and (LiH)\(_{32}\) systems each require approximately 32, 80, and 235 MB of RAM per core, respectively, across 512 cores. The Si\(_{8}\),  Si\(_{32}\), and Si\(_{64}\) systems each require approximately 42, 233, and 836 MB of RAM per core, respectively, across 512 cores.

Solving Sternhiemer equations occupies a large majority of the runtime when computing the RPA correlation energy. In particular, the runtime for solving the Sternheimer equations occupies \(63\%\), \(89\%\), and \(98\%\) of the total runtime on 1,024 cores for (LiH)\(_4\), (LiH)\(_{16}\), and (LiH)\(_{32}\). The overall runtime for solving the RPA correlation energy is dependent on the number of iterations taken to solve the Sternheimer equations, which varies strongly with \(\omega_k\), as seen in Tables \ref{table:si_iterations_comparison} and \ref{table:lih_iterations_comparison}. As \(\omega_k\) decreases, more iterations are required for convergence alongside having much more variance. This is likely because \(\omega_k\) values close to 0 lead to ill-conditioning in the Sternheimer linear system, as observed by Shah et al.\ \cite{SHAH_DISSERTATION}.

We compare the CPU time taken to compute the RPA correlation energy for (LiH)\(_4\) between SPARC and ABINIT. Zhang et al.\ performed ABINIT performance tests on the same HPC system as our experiments \cite{ZHANG_SQ}. Zhang et al.\ utilized a planewave cutoff of 75 ha, which converged the RPA correlation energy to an error of \(~0.001\) Ha/atom, which is comparable to the accuracy of the proposed method. Zhang el al.\ note that the CPU time taken by ABINIT on one core is 34 hours. However, our Lanczos quadrature-based implementation in SPARC takes 0.34 hours on one core, indicating that SPARC obtains better speedups over ABINIT.

\begin{table}[H]
\centering
\begin{tabular}{|c|c|c|c|c|c|c|c|c|c|}
\hline
\multirow{2}{*}{\textbf{$\omega$}} & \multicolumn{3}{c|}{\textbf{Mean}} & \multicolumn{3}{c|}{\textbf{Std. Dev.}} & \multicolumn{3}{c|}{\textbf{(Min, Max)}} \\
\cline{2-10}
 & \textbf{Si\(_8\)} & \textbf{Si\(_{32}\)} & \textbf{Si\(_{64}\)} & \textbf{Si\(_8\)} & \textbf{Si\(_{32}\)} & \textbf{Si\(_{64}\)} & \textbf{Si\(_8\)} & \textbf{Si\(_{32}\)} & \textbf{Si\(_{64}\)} \\
\hline
49.37 & 1.31 & 1.30 & 1.31 & 0.46 & 0.46 & 0.46 & (1, 2) & (1, 2) & (1, 2) \\
8.84  & 3.14 & 2.88 & 2.87 & 0.99 & 1.14 & 1.14 & (1, 4) & (1, 4) & (1, 4) \\
3.22  & 5.72 & 5.25 & 5.25 & 1.03 & 1.37 & 1.36 & (2, 7) & (1, 7) & (1, 7) \\
1.45  & 8.53 & 7.61 & 7.63 & 1.18 & 1.70 & 1.69 & (4, 11) & (1, 11) & (1, 11) \\
0.69  & 12.20 & 10.08 & 10.19 & 1.96 & 2.51 & 2.42 & (4, 17) & (1, 16) & (2, 16) \\
0.31  & 16.19 & 12.38 & 12.54 & 3.88 & 4.15 & 3.99 & (5, 24) & (0, 24) & (1, 23) \\
0.11  & 20.75 & 14.42 & 14.32 & 7.96 & 6.61 & 6.27 & (4, 42) & (0, 38) & (0, 36) \\
0.02  & 27.34 & 15.37 & 14.82 & 17.20 & 8.62 & 7.66 & (4, 90) & (0, 79) & (0, 47) \\
\hline
\end{tabular}
\caption{Iterations required to solve Sternheimer equations for each grid point for a given \(\omega_k\) for Si strong scaling systems.}
\label{table:si_iterations_comparison}
\end{table}

\begin{table}[H]
\centering
\begin{tabular}{|c|c|c|c|c|c|c|c|c|c|}
\hline
\multirow{2}{*}{\textbf{$\omega$}} & \multicolumn{3}{c|}{\textbf{Mean}} & \multicolumn{3}{c|}{\textbf{Std. Dev.}} & \multicolumn{3}{c|}{\textbf{(Min, Max)}} \\
\cline{2-10}
 & \footnotesize{\textbf{(LiH)\(_4\)}} & \footnotesize{\textbf{(LiH)\(_{16}\)}} & \footnotesize{\textbf{(LiH)\(_{32}\)}} & \footnotesize{\textbf{(LiH)\(_4\)}} & \footnotesize{\textbf{(LiH)\(_{16}\)}} & \footnotesize{\textbf{(LiH)\(_{32}\)}} & \footnotesize{\textbf{(LiH)\(_4\)}} & \footnotesize{\textbf{(LiH)\(_{16}\)}} & \footnotesize{\textbf{(LiH)\(_{32}\)}} \\
\hline
49.37 & 1.31 & 1.12 & 1.10 & 0.52 & 0.63 & 0.64 & (0, 2) & (0, 2) & (0, 2) \\
8.84  & 3.13 & 2.53 & 2.44 & 0.98 & 1.41 & 1.44 & (0, 5) & (0, 5) & (0, 5) \\
3.22  & 5.26 & 4.33 & 4.22 & 1.75 & 2.50 & 2.56 & (0, 8) & (0, 8) & (0, 8) \\
1.45  & 7.09 & 5.96 & 5.86 & 2.94 & 3.84 & 3.91 & (0, 11) & (0, 11) & (0, 11) \\
0.69  & 8.72 & 7.46 & 7.45 & 4.43 & 5.33 & 5.46 & (0, 17) & (0, 16) & (0, 16) \\
0.31  & 10.38 & 8.93 & 8.96 & 6.20 & 6.88 & 6.98 & (0, 23) & (0, 21) & (0, 21) \\
0.11  & 11.60 & 9.80 & 9.83 & 7.77 & 7.87 & 7.90 & (0, 29) & (0, 25) & (0, 26) \\
0.02  & 12.02 & 10.00 & 10.03 & 8.41 & 8.11 & 8.12 & (0, 31) & (0, 26) & (0, 28) \\
\hline
\end{tabular}
\caption{Iterations required to solve Sternheimer equations for each grid point for a given \(\omega_k\) for LiH strong scaling systems.}
\label{table:lih_iterations_comparison}
\end{table}

\section{Conclusions and Future Work}

We presented a highly scalable framework for computing the RPA correlation energy in real-space, which we implemented within the open-source SPARC code and which achieves near-cubic scaling. We showed it is possible to exploit spatial smoothness of the correlation energy density by computing a portion of the diagonal entries of a matrix function that lie on a coarse grid, then approximating the fine grid values via an interpolation method, such as cubic or Fourier interpolation, to significantly reduce the computational prefactor by up to a factor of \(8\).
Furthermore, we systematically analyzed the role of error cancellation in the global sum of energy densities.
We also achieved near ideal scaling on a system with 256 valence electrons on up to 4,096 CPU cores.

Possible avenues for future work include supporting non-\(\Gamma\)-point Brillouin zone integration, as this would increase the range of chemical systems that are accessible for the proposed RPA calculation method. Future work also includes using block Lanczos quadrature, which compute the energies for multiple grid points simultaneously via block operations. Developing a multi-GPU implementation of the proposed framework is also worthwhile, as such block operations can be accelerated on tensor cores \cite{10.1145/3815589,CLARK201829}. Additionally, incorporating preconditioning may accelerate the solution to the Sternheimer equations. A possible direction for future work also includes exploring advanced machine learning-driven interpolation strategies that rely on a sparse set of sample points or significantly coarser grids. Such methods, if successful, can reduce the prefactor of the proposed method beyond factors of \(8\).
\section*{Data and Software Availability Statement}
The data and source code underlying this study are available in the published article and its Supporting Information. 

\section*{Acknowledgments}
The authors acknowledge the support of the U.S. Department of Energy, Office of Science under grant DE-SC0023445. This work was performed at the HPC facilities operated by Partnership for an Advanced Computing Environment (PACE) through its Phoenix cluster at Georgia Institute of Technology in Atlanta, Georgia. This work was performed in part under the auspices of the U.S. DOE by Lawrence Livermore National Laboratory (LLNL) under Contract DE-AC52-07NA27344. This material is based upon work supported by the U.S. Department
of Energy, Office of Science, and Office of Advanced Scientific Computing Research under Award Number DE-SC0026617. 

\section*{Supporting Information}
The Supporting Information, provided as a ZIP archive, includes the source code, input files, and output files required to reproduce the calculations reported in this work.

\printbibliography

@ARTICLE{Kohn1965-sf,
  title     = "Self-consistent equations including exchange and correlation
               effects",
  author    = "Kohn, W and Sham, L J",
  journal   = "Phys. Rev.",
  publisher = "American Physical Society (APS)",
  volume    =  140,
  number    = "4A",
  pages     = "A1133--A1138",
  month     =  nov,
  year      =  1965,
  copyright = "http://link.aps.org/licenses/aps-default-license"
}

@ARTICLE{Hohenberg1964-qx,
  title     = "Inhomogeneous Electron Gas",
  author    = "Hohenberg, P and Kohn, W",
  journal   = "Phys. Rev.",
  publisher = "American Physical Society (APS)",
  volume    =  136,
  number    = "3B",
  pages     = "B864--B871",
  month     =  nov,
  year      =  1964,
  copyright = "http://link.aps.org/licenses/aps-default-license"
}

@INPROCEEDINGS{Perdew2001-lw,
  title      = "Jacob's ladder of density functional approximations for the
                exchange-correlation energy",
  booktitle  = "{AIP} Conference Proceedings",
  author     = "Perdew, John P",
  publisher  = "AIP",
  year       =  2001,
  conference = "Density functional theory and its application to materials",
  location   = "Antwerp (Belgium)"
}

@ARTICLE{Ren2012-vl,
  title     = "Random-phase approximation and its applications in computational
               chemistry and materials science",
  author    = "Ren, Xinguo and Rinke, Patrick and Joas, Christian and
               Scheffler, Matthias",
  journal   = "J. Mater. Sci.",
  publisher = "Springer Science and Business Media LLC",
  volume    =  47,
  number    =  21,
  pages     = "7447--7471",
  month     =  nov,
  year      =  2012,
  language  = "en"
}

@ARTICLE{Eshuis2012-lg,
  title     = "Electron correlation methods based on the random phase
               approximation",
  author    = "Eshuis, Henk and Bates, Jefferson E and Furche, Filipp",
  journal   = "Theor. Chem. Acc.",
  publisher = "Springer Science and Business Media LLC",
  volume    =  131,
  number    =  1,
  month     =  jan,
  year      =  2012,
  language  = "en"
}

@ARTICLE{Gonze2016-pr,
  title     = "Recent developments in the {ABINIT} software package",
  author    = "Gonze, X and Jollet, F and Abreu Araujo, F and Adams, D and
               Amadon, B and Applencourt, T and Audouze, C and Beuken, J-M and
               Bieder, J and Bokhanchuk, A and Bousquet, E and Bruneval, F and
               Caliste, D and C{\^o}t{\'e}, M and Dahm, F and Da Pieve, F and
               Delaveau, M and Di Gennaro, M and Dorado, B and Espejo, C and
               Geneste, G and Genovese, L and Gerossier, A and Giantomassi, M
               and Gillet, Y and Hamann, D R and He, L and Jomard, G and
               Laflamme Janssen, J and Le Roux, S and Levitt, A and Lherbier, A
               and Liu, F and Luka{\v c}evi{\'c}, I and Martin, A and Martins,
               C and Oliveira, M J T and Ponc{\'e}, S and Pouillon, Y and
               Rangel, T and Rignanese, G-M and Romero, A H and Rousseau, B and
               Rubel, O and Shukri, A A and Stankovski, M and Torrent, M and
               Van Setten, M J and Van Troeye, B and Verstraete, M J and
               Waroquiers, D and Wiktor, J and Xu, B and Zhou, A and Zwanziger,
               J W",
  journal   = "Comput. Phys. Commun.",
  publisher = "Elsevier BV",
  volume    =  205,
  pages     = "106--131",
  month     =  aug,
  year      =  2016,
  language  = "en"
}

@ARTICLE{Enkovaara2010-xo,
  title     = "Electronic structure calculations with {GPAW}: a real-space
               implementation of the projector augmented-wave method",
  author    = "Enkovaara, J and Rostgaard, C and Mortensen, J J and Chen, J and
               Du{\l}ak, M and Ferrighi, L and Gavnholt, J and Glinsvad, C and
               Haikola, V and Hansen, H A and Kristoffersen, H H and Kuisma, M
               and Larsen, A H and Lehtovaara, L and Ljungberg, M and
               Lopez-Acevedo, O and Moses, P G and Ojanen, J and Olsen, T and
               Petzold, V and Romero, N A and Stausholm-M{\o}ller, J and
               Strange, M and Tritsaris, G A and Vanin, M and Walter, M and
               Hammer, B and H{\"a}kkinen, H and Madsen, G K H and Nieminen, R
               M and N{\o}rskov, J K and Puska, M and Rantala, T T and
               Schi{\o}tz, J and Thygesen, K S and Jacobsen, K W",
  journal   = "J. Phys. Condens. Matter",
  publisher = "IOP Publishing",
  volume    =  22,
  number    =  25,
  pages     = "253202",
  month     =  jun,
  year      =  2010,
  language  = "en"
}

@ARTICLE{Kresse1996-oo,
  title     = "Efficient iterative schemes for ab initio total-energy
               calculations using a plane-wave basis set",
  author    = "Kresse, G and Furthm{\"u}ller, J",
  journal   = "Phys. Rev. B Condens. Matter",
  publisher = "American Physical Society (APS)",
  volume    =  54,
  number    =  16,
  pages     = "11169--11186",
  month     =  oct,
  year      =  1996,
  copyright = "http://link.aps.org/licenses/aps-default-license",
  language  = "en"
}

@ARTICLE{Rocca2014-wj,
  title     = "Random-phase approximation correlation energies from Lanczos
               chains and an optimal basis set: theory and applications to the
               benzene dimer",
  author    = "Rocca, Dario",
  journal   = "J. Chem. Phys.",
  publisher = "AIP Publishing",
  volume    =  140,
  number    =  18,
  pages     = "18A501",
  month     =  may,
  year      =  2014,
  language  = "en"
}

@ARTICLE{Del_Ben2013-kg,
  title     = "Electron correlation in the condensed phase from a resolution of
               identity approach based on the Gaussian and Plane Waves scheme",
  author    = "Del Ben, Mauro and Hutter, J{\"u}rg and VandeVondele, Joost",
  journal   = "J. Chem. Theory Comput.",
  publisher = "American Chemical Society (ACS)",
  volume    =  9,
  number    =  6,
  pages     = "2654--2671",
  month     =  jun,
  year      =  2013,
  language  = "en"
}

@ARTICLE{Kaltak2014-ln,
  title     = "Low scaling algorithms for the random phase approximation:
               Imaginary time and Laplace transformations",
  author    = "Kaltak, Merzuk and Klime{\v s}, Ji{\v r}{\'\i} and Kresse, Georg",
  journal   = "J. Chem. Theory Comput.",
  publisher = "American Chemical Society (ACS)",
  volume    =  10,
  number    =  6,
  pages     = "2498--2507",
  month     =  jun,
  year      =  2014,
  language  = "en"
}

@ARTICLE{Nguyen2014-av,
  title     = "Ab initioself-consistent total-energy calculations within the
               {EXX/RPA} formalism",
  author    = "Nguyen, Ngoc Linh and Colonna, Nicola and de Gironcoli, Stefano",
  journal   = "Phys. Rev. B Condens. Matter Mater. Phys.",
  publisher = "American Physical Society (APS)",
  volume    =  90,
  number    =  4,
  month     =  jul,
  year      =  2014,
  copyright = "http://link.aps.org/licenses/aps-default-license"
}

@ARTICLE{Weinberg2024-ru,
  title     = "Static subspace approximation for random phase approximation
               correlation energies: Implementation and performance",
  author    = "Weinberg, Daniel and Hull, Olivia A and Clary, Jacob M and
               Sundararaman, Ravishankar and Vigil-Fowler, Derek and Del Ben,
               Mauro",
  journal   = "J. Chem. Theory Comput.",
  publisher = "American Chemical Society (ACS)",
  month     =  sep,
  year      =  2024,
  language  = "en"
}

@ARTICLE{Lu2017-wq,
  title     = "Cubic scaling algorithms for {RPA} correlation using
               interpolative separable density fitting",
  author    = "Lu, Jianfeng and Thicke, Kyle",
  journal   = "J. Comput. Phys.",
  publisher = "Elsevier BV",
  volume    =  351,
  pages     = "187--202",
  month     =  dec,
  year      =  2017,
  language  = "en"
}

@ARTICLE{Del_Ben2015-kn,
  title     = "Enabling simulation at the fifth rung of {DFT}: Large scale
               {RPA} calculations with excellent time to solution",
  author    = "Del Ben, Mauro and Sch{\"u}tt, Ole and Wentz, Tim and Messmer,
               Peter and Hutter, J{\"u}rg and VandeVondele, Joost",
  journal   = "Comput. Phys. Commun.",
  publisher = "Elsevier BV",
  volume    =  187,
  pages     = "120--129",
  month     =  feb,
  year      =  2015,
  language  = "en"
}

@ARTICLE{Ghosh2017-mc,
  title     = "{SPARC}: Accurate and efficient finite-difference formulation
               and parallel implementation of Density Functional Theory:
               Isolated clusters",
  author    = "Ghosh, Swarnava and Suryanarayana, Phanish",
  journal   = "Comput. Phys. Commun.",
  publisher = "Elsevier BV",
  volume    =  212,
  pages     = "189--204",
  month     =  mar,
  year      =  2017,
  language  = "en"
}

@ARTICLE{Xu2021-he,
  title     = "{SPARC}: Simulation package for ab-initio real-space
               calculations",
  author    = "Xu, Qimen and Sharma, Abhiraj and Comer, Benjamin and Huang, Hua
               and Chow, Edmond and Medford, Andrew J and Pask, John E and
               Suryanarayana, Phanish",
  journal   = "SoftwareX",
  publisher = "Elsevier BV",
  volume    =  15,
  number    =  100709,
  pages     = "100709",
  month     =  jul,
  year      =  2021,
  copyright = "http://creativecommons.org/licenses/by-nc-nd/4.0/",
  language  = "en"
}

@ARTICLE{Zhang2024-vi,
  title     = "{SPARC} v2.0.0: Spin-orbit coupling, dispersion interactions,
               and advanced exchange--correlation functionals",
  author    = "Zhang, Boqin and Jing, Xin and Xu, Qimen and Kumar, Shashikant
               and Sharma, Abhiraj and Erlandson, Lucas and Sahoo, Sushree
               Jagriti and Chow, Edmond and Medford, Andrew J and Pask, John E
               and Suryanarayana, Phanish",
  journal   = "Softw. Impacts",
  publisher = "Elsevier BV",
  volume    =  20,
  number    =  100649,
  pages     = "100649",
  month     =  may,
  year      =  2024,
  copyright = "http://creativecommons.org/licenses/by-nc-nd/4.0/",
  language  = "en"
}

@ARTICLE{Jing2024-yn,
  title     = "Efficient real space formalism for hybrid density functionals",
  author    = "Jing, Xin and Suryanarayana, Phanish",
  journal   = "J. Chem. Phys.",
  publisher = "AIP Publishing",
  volume    =  161,
  number    =  8,
  pages     = "084115",
  month     =  aug,
  year      =  2024,
  language  = "en"
}

@ARTICLE{Jing2025-ei,
  title    = "{GPU} acceleration of hybrid functional calculations in the
              {SPARC} electronic structure code",
  author   = "Jing, Xin and Sharma, Abhiraj and Pask, John E and Suryanarayana,
              Phanish",
  journal  = "J. Chem. Phys.",
  volume   =  162,
  number   =  18,
  pages    = "184105",
  month    =  may,
  year     =  2025,
  language = "en"
}

@ARTICLE{Van_der_Vorst1990-dd,
  title     = "A {Petrov-Galerkin} type method for solving Axk=b, where A is
               symmetric complex",
  author    = "van der Vorst, H A and Melissen, J B M",
  journal   = "IEEE Trans. Magn.",
  publisher = "Institute of Electrical and Electronics Engineers (IEEE)",
  volume    =  26,
  number    =  2,
  pages     = "706--708",
  month     =  mar,
  year      =  1990,
  copyright = "https://ieeexplore.ieee.org/Xplorehelp/downloads/license-information/IEEE.html"
}

@INPROCEEDINGS{SC_RPA,
  author={Shah, Shikhar and Zhang, Boqin and Huang, Hua and Pask, John E. and Suryanarayana, Phanish and Chow, Edmond},
  booktitle={SC24: International Conference for High Performance Computing, Networking, Storage and Analysis}, 
  title={Many-Body Electronic Correlation Energy using Krylov Subspace Linear Solvers}, 
  year={2024},
  volume={},
  number={},
  pages={1-15},
  doi={10.1109/SC41406.2024.00066}}

@phdthesis{SHAH_DISSERTATION,
author={Shah,Shikhar},
year={2024},
title={Block Iterative Methods with Applications to Density Functional Theory},
school       = {Georgia Institute of Technology},
pages={136},
url={https://repository.gatech.edu/server/api/core/bitstreams/ca234013-778b-4244-a072-12bb4255e5df/content},
}

@article{ZHANG_SQ,
author = {Zhang, Boqin and Shah, Shikhar and Pask, John E. and Chow, Edmond and Suryanarayana, Phanish},
title = {Random Phase Approximation Correlation Energy Using Real-Space Density Functional Perturbation Theory},
journal = {Journal of Chemical Theory and Computation},
volume = {21},
number = {12},
pages = {6023-6033},
year = {2025},
doi = {10.1021/acs.jctc.5c00528},
    note ={PMID: 40503615},

URL = { 
    
        https://doi.org/10.1021/acs.jctc.5c00528
    
    

},
eprint = { 
    
        https://doi.org/10.1021/acs.jctc.5c00528
    
    

}

}

@article{BerkeleyGW,
title = {BerkeleyGW: A massively parallel computer package for the calculation of the quasiparticle and optical properties of materials and nanostructures},
journal = {Computer Physics Communications},
volume = {183},
number = {6},
pages = {1269-1289},
year = {2012},
issn = {0010-4655},
doi = {https://doi.org/10.1016/j.cpc.2011.12.006},
url = {https://www.sciencedirect.com/science/article/pii/S0010465511003912},
author = {Jack Deslippe and Georgy Samsonidze and David A. Strubbe and Manish Jain and Marvin L. Cohen and Steven G. Louie}
}

@ARTICLE{WEST,
  title     = "Large scale {GW} calculations",
  author    = "Govoni, Marco and Galli, Giulia",
  journal   = "J. Chem. Theory Comput.",
  publisher = "American Chemical Society (ACS)",
  volume    =  11,
  number    =  6,
  pages     = "2680--2696",
  month     =  jun,
  year      =  2015,
  copyright = "http://pubs.acs.org/page/policy/authorchoice\_termsofuse.html",
  language  = "en"
}

@INPROCEEDINGS{9835388,
  author={Ayala, Alan and Tomov, Stan and Stoyanov, Miroslav and Haidar, Azzam and Dongarra, Jack},
  booktitle={2022 IEEE International Parallel and Distributed Processing Symposium Workshops (IPDPSW)}, 
  title={Performance Analysis of Parallel FFT on Large Multi-GPU Systems}, 
  year={2022},
  volume={},
  number={},
  pages={372-381},
  doi={10.1109/IPDPSW55747.2022.00072}}

@INCOLLECTION{Ayala2021-wj,
  title     = "Scalability Issues in {FFT} Computation",
  booktitle = "Lecture Notes in Computer Science",
  author    = "Ayala, Alan and Tomov, Stanimire and Stoyanov, Miroslav and
               Dongarra, Jack",
  publisher = "Springer International Publishing",
  pages     = "279--287",
  series    = "Lecture Notes in Computer Science",
  year      =  2021,
  address   = "Cham"
}

@INPROCEEDINGS{9652837,
  author={Ayala, Alan and Tomov, Stan and Stoyanov, Miroslav and Haidar, Azzam and Dongarra, Jack},
  booktitle={2021 Workshop on Exascale MPI (ExaMPI)}, 
  title={Accelerating Multi - Process Communication for Parallel 3-D FFT}, 
  year={2021},
  volume={},
  number={},
  pages={46-53},
  doi={10.1109/ExaMPI54564.2021.00011}}

@ARTICLE{Schlipf2020-zc,
  title     = "{SternheimerGW}: A program for calculating {GW} quasiparticle
               band structures and spectral functions without unoccupied states",
  author    = "Schlipf, Martin and Lambert, Henry and Zibouche, Nourdine and
               Giustino, Feliciano",
  journal   = "Comput. Phys. Commun.",
  publisher = "Elsevier BV",
  volume    =  247,
  number    =  106856,
  pages     = "106856",
  month     =  feb,
  year      =  2020,
  language  = "en"
}

@BOOK{Goncharov2014-sk,
  title     = "Non-linear optical response in atoms, molecules and clusters; An
               explicit time dependent density functional approach",
  author    = "Goncharov, Vladimir",
  publisher = "Springer International Publishing",
  series    = "SpringerBriefs in Molecular Science",
  edition   =  2014,
  month     =  jan,
  year      =  2014,
  copyright = "https://www.springernature.com/gp/researchers/text-and-data-mining",
  language  = "en"
}

@BOOK{Martin2020-em,
  title     = "Electronic structure",
  author    = "Martin, Richard M",
  publisher = "Cambridge University Press",
  edition   =  2,
  month     =  aug,
  year      =  2020,
  address   = "Cambridge, England",
  language  = "en"
}

@ARTICLE{Saad2010-kc,
  title     = "Numerical methods for electronic structure calculations of
               materials",
  author    = "Saad, Yousef and Chelikowsky, James R and Shontz, Suzanne M",
  journal   = "SIAM Rev. Soc. Ind. Appl. Math.",
  publisher = "Society for Industrial \& Applied Mathematics (SIAM)",
  volume    =  52,
  number    =  1,
  pages     = "3--54",
  month     =  jan,
  year      =  2010
}

@article{10.1145/3815589,
author = {Li, Wen and Zhang, Zheng and Zhao, Jie and Zhu, Song and Hui, Ming},
title = {BLR-Krylov: A Single-GPU Iterative SpMM Framework with Communication Avoidance and Block Low-Rank Optimization},
year = {2026},
issue_date = {June 2026},
publisher = {Association for Computing Machinery},
address = {New York, NY, USA},
volume = {23},
number = {2},
issn = {1544-3566},
url = {https://doi.org/10.1145/3815589},
doi = {10.1145/3815589},
journal = {ACM Trans. Archit. Code Optim.},
month = jun,
articleno = {68},
numpages = {32}
}

@article{CLARK201829,
title = {Pushing memory bandwidth limitations through efficient implementations of Block-Krylov space solvers on GPUs},
journal = {Computer Physics Communications},
volume = {233},
pages = {29-40},
year = {2018},
issn = {0010-4655},
doi = {https://doi.org/10.1016/j.cpc.2018.06.019},
url = {https://www.sciencedirect.com/science/article/pii/S0010465518302273},
author = {M.A. Clark and Alexei Strelchenko and Alejandro Vaquero and Mathias Wagner and Evan Weinberg}
}

@article{PhysRevLett.77.3865,
  title = {Generalized Gradient Approximation Made Simple},
  author = {Perdew, John P. and Burke, Kieron and Ernzerhof, Matthias},
  journal = {Phys. Rev. Lett.},
  volume = {77},
  issue = {18},
  pages = {3865--3868},
  numpages = {0},
  year = {1996},
  month = {Oct},
  publisher = {American Physical Society},
  doi = {10.1103/PhysRevLett.77.3865},
  url = {https://link.aps.org/doi/10.1103/PhysRevLett.77.3865}
}

@article{PhysRevB.88.085117,
  title = {Optimized norm-conserving Vanderbilt pseudopotentials},
  author = {Hamann, D. R.},
  journal = {Phys. Rev. B},
  volume = {88},
  issue = {8},
  pages = {085117},
  numpages = {10},
  year = {2013},
  month = {Aug},
  publisher = {American Physical Society},
  doi = {10.1103/PhysRevB.88.085117},
  url = {https://link.aps.org/doi/10.1103/PhysRevB.88.085117}
}

@article{SHOJAEI2023108594,
title = {Soft and transferable pseudopotentials from multi-objective optimization},
journal = {Computer Physics Communications},
volume = {283},
pages = {108594},
year = {2023},
issn = {0010-4655},
doi = {https://doi.org/10.1016/j.cpc.2022.108594},
url = {https://www.sciencedirect.com/science/article/pii/S0010465522003137},
author = {Mostafa Faghih Shojaei and John E. Pask and Andrew J. Medford and Phanish Suryanarayana}
}

@BOOK{laug,
      AUTHOR = {Anderson, E. and Bai, Z. and Bischof, C. and
                Blackford, S. and Demmel, J. and Dongarra, J. and
                Du Croz, J. and Greenbaum, A. and Hammarling, S. and
                McKenney, A. and Sorensen, D.},
      TITLE = {{LAPACK} Users' Guide},
      EDITION = {Third},
      PUBLISHER = {Society for Industrial and Applied Mathematics},
      YEAR = {1999},
      ADDRESS = {Philadelphia, PA},
      ISBN = {0-89871-447-8 (paperback)} }

@inproceedings{gabriel2004open,
  title={Open {MPI}: Goals, concept, and design of a next generation {MPI} implementation},
  author={Gabriel, Edgar and Fagg, Graham E and Bosilca, George and Angskun, Thara and Dongarra, Jack J and Squyres, Jeffrey M and Sahay, Vishal and Kambadur, Prabhanjan and Barrett, Brian and Lumsdaine, Andrew and others},
  booktitle={Recent Advances in Parallel Virtual Machine and Message Passing Interface: 11th European PVM/MPI Users' Group Meeting, Budapest, Hungary, September 19-22, 2004. Proceedings 11},
  pages={126--140},
  year={2004},
  organization={Springer}
}

@article{doi:10.1137/16M1104974,
author = {Ubaru, Shashanka and Chen, Jie and Saad, Yousef},
title = {Fast Estimation of \$tr(f(A))\$ via Stochastic Lanczos Quadrature},
journal = {SIAM Journal on Matrix Analysis and Applications},
volume = {38},
number = {4},
pages = {1075-1099},
year = {2017},
doi = {10.1137/16M1104974},

URL = { 
    
        https://doi.org/10.1137/16M1104974
    
    

},
eprint = { 
    
        https://doi.org/10.1137/16M1104974
    
    

}
}

@article{10.1021/acs.jctc.1c01291,
    author = {Poier, Pier Paolo and Lagardère, Louis and Piquemal, Jean-Philip},
    title = {O(N) Stochastic Evaluation
of Many-Body van
der Waals Energies in Large Complex Systems},
    journal = {Journal of Chemical Theory and Computation},
    volume = {18},
    number = {3},
    pages = {1633-1645},
    year = {2022},
    month = {02},
    issn = {1549-9618},
    doi = {10.1021/acs.jctc.1c01291},
    url = {https://doi.org/10.1021/acs.jctc.1c01291},
    eprint = {https://pubs.acs.org/jctcce/article-pdf/18/3/1633/2252/ct1c01291.pdf},
}

@ARTICLE{White2021-dg,
  title         = "Quantum harmonic free energies for biomolecules and
                   nanomaterials",
  author        = "White, Alec F and Li, Chenghan and Zhang, Xing and Chan,
                   Garnet Kin-Lic",
  month         =  nov,
  year          =  2021,
  copyright     = "http://creativecommons.org/licenses/by/4.0/",
  archivePrefix = "arXiv",
  primaryClass  = "physics.chem-ph",
  eprint        = "2111.12200"
}

@article{PhysRevA.68.032507,
  title = {Exchange-correlation potentials in the adiabatic connection fluctuation-dissipation framework},
  author = {Niquet, Y. M. and Fuchs, M. and Gonze, X.},
  journal = {Phys. Rev. A},
  volume = {68},
  issue = {3},
  pages = {032507},
  numpages = {13},
  year = {2003},
  month = {Sep},
  publisher = {American Physical Society},
  doi = {10.1103/PhysRevA.68.032507},
  url = {https://link.aps.org/doi/10.1103/PhysRevA.68.032507}
}

@article{PhysRev.126.413,
  title = {Quantum Theory of the Dielectric Constant in Real Solids},
  author = {Adler, Stephen L.},
  journal = {Phys. Rev.},
  volume = {126},
  issue = {2},
  pages = {413--420},
  numpages = {0},
  year = {1962},
  month = {Apr},
  publisher = {American Physical Society},
  doi = {10.1103/PhysRev.126.413},
  url = {https://link.aps.org/doi/10.1103/PhysRev.126.413}
}

@article{PhysRev.129.62,
  title = {Dielectric Constant with Local Field Effects Included},
  author = {Wiser, Nathan},
  journal = {Phys. Rev.},
  volume = {129},
  issue = {1},
  pages = {62--69},
  numpages = {0},
  year = {1963},
  month = {Jan},
  publisher = {American Physical Society},
  doi = {10.1103/PhysRev.129.62},
  url = {https://link.aps.org/doi/10.1103/PhysRev.129.62}
}

@article{kronik2006parsec,
  title={PARSEC--the pseudopotential algorithm for real-space electronic structure calculations: recent advances and novel applications to nano-structures},
  author={Kronik, Leeor and Makmal, Adi and Tiago, Murilo L and Alemany, MMG and Jain, Manish and Huang, Xiangyang and Saad, Yousef and Chelikowsky, James R},
  journal={physica status solidi (b)},
  volume={243},
  number={5},
  pages={1063--1079},
  year={2006},
  publisher={Wiley Online Library}
}

@article{castro2006octopus,
  title={Octopus: a tool for the application of time-dependent density functional theory},
  author={Castro, Alberto and Appel, Heiko and Oliveira, Micael and Rozzi, Carlo A and Andrade, Xavier and Lorenzen, Florian and Marques, Miguel AL and Gross, EKU and Rubio, Angel},
  journal={physica status solidi (b)},
  volume={243},
  number={11},
  pages={2465--2488},
  year={2006},
  publisher={Wiley Online Library}
}

@article{motamarri2020dft,
  title={DFT-FE--A massively parallel adaptive finite-element code for large-scale density functional theory calculations},
  author={Motamarri, Phani and Das, Sambit and Rudraraju, Shiva and Ghosh, Krishnendu and Davydov, Denis and Gavini, Vikram},
  journal={Computer Physics Communications},
  volume={246},
  pages={106853},
  year={2020},
  publisher={Elsevier}
}

@article{beck2009real,
  title={Real-space and multigrid methods in computational chemistry},
  author={Beck, Thomas L},
  journal={REVIEWS IN COMPUTATIONAL CHEMISTRY, VOL 26},
  volume={26},
  pages={223--285},
  year={2009},
  publisher={JOHN WILEY \& SONS INC 111 RIVER ST, HOBOKEN, NJ 07030 USA}
}

@article{bhowmik2026bulk,
title={Bulk Boundary Condition for Surface Calculations in Density Functional Theory},
author={Bhowmik, Sayan and Medford, Andrew J and Suryanarayana, Phanish},
journal={arXiv preprint arXiv:2607.07894},
year={2026}
}


\end{document}